\documentclass[acmsmall,screen]{acmart}

\usepackage{algorithm,algpseudocode}

\usepackage{graphicx} % Required for inserting images
\usepackage{graphics}
\usepackage{xspace}
\usepackage{booktabs,makecell,tabularx}

\usepackage{multirow}

\usepackage{colortbl}
\usepackage{tcolorbox}
\usepackage{xcolor}
\usepackage{enumitem}
\usepackage{pifont}
\usepackage{amsmath,amsfonts}
\usepackage{float}
\usepackage{subfig}

\usepackage{array}
\usepackage{subfig}
\usepackage{textcomp}
\usepackage{stfloats}
\usepackage{url}
\usepackage{verbatim}
\usepackage{graphicx}
\usepackage{fontawesome}

\usepackage{mdframed}

\usepackage{tcolorbox}

\definecolor{zirui}{RGB}{16,109,156}

\newcommand{\appname}{{\sc Lumen}\xspace}
\newcommand{\baselinename}{{\sc Rvm}\xspace}
\newcommand{\baselinenamebold}{{\sc \textbf{Rvm}}\xspace}
\newcommand{\appnamebold}{{\sc \textbf{Lumen}}\xspace}
\newcommand{\think}{{\sc ThinkRepair}\xspace}
\newcommand{\thinkbold}{{\sc \textbf{ThinkRepair}}\xspace}
\newcommand{\master}{{\sc VulMaster}\xspace}

\newcommand{\mussel}{{\sc MUSSEL}\xspace}
\newcommand{\musselbold}{{\sc \textbf{MUSSEL }}\xspace}

\newcommand{\VulMaster}{{\sc VulMaster}\xspace}
\newcommand{\VulMasterbold}{{\sc \textbf{VulMaster}}\xspace}

\definecolor{myyellow}{HTML}{FFF2CC}
\definecolor{myblue}{RGB}{255,255,255}

\newcommand{\dx}[1]{\textcolor{black}{{#1}}}

\newcommand{\xx}[1]{\textcolor{black}{{#1}}}

\AtBeginDocument{%
  }

\setcopyright{acmlicensed}
\copyrightyear{2018}
\acmYear{2018}
\acmDOI{XXXXXXX.XXXXXXX}
\acmConference[Conference acronym 'XX]{Make sure to enter the correct
  conference title from your rights confirmation email}{June 03--05,
  2018}{Woodstock, NY}
\acmISBN{978-1-4503-XXXX-X/2018/06}

\begin{document}

\title{Generating Mitigations for Downstream Projects to Neutralize Upstream Library Vulnerability }

\author{Zirui Chen}
\orcid{0009-0004-6236-9150}
\affiliation{%
  \institution{The State Key Laboratory of Blockchain and Data Security, Zhejiang University}
   \city{Hangzhou}
  \country{China}
}
\email{chenzirui@zju.edu.cn}

\author{Puhua Sun}
\orcid{0009-0000-0375-3241}
\affiliation{%
  \institution{The State Key Laboratory of Blockchain and Data Security, Zhejiang University}
   \city{Hangzhou}
  \country{China}
}
\email{puhuasun@zju.edu.cn}

\author{Xing Hu}
\orcid{0000-0003-0093-3292}
\authornote{Corresponding author.}
\affiliation{
  \institution{The State Key Laboratory of Blockchain and Data Security, Zhejiang University}
   \city{Hangzhou}
  \country{China}
}
\email{xinghu@zju.edu.cn}

\author{Xin Xia}
\orcid{0000-0002-6302-3256}
\affiliation{%
  \institution{The State Key Laboratory of Blockchain and Data Security, Zhejiang University}
  \city{Hangzhou}
  \country{China}
}
\affiliation{%
  \institution{Hangzhou High-Tech Zone (Binjiang) Institute of Blockchain and Data Security}
  \city{Hangzhou}
  \country{China}
}
\email{xin.xia@acm.org}

\author{Xiaohu Yang}
\orcid{0000-0003-4111-4189}
\affiliation{%
  \institution{The State Key Laboratory of Blockchain and Data Security, Zhejiang University}
   \city{Hangzhou}
  \country{China}
}
\email{yangxh@zju.edu.cn}

\renewcommand{\shortauthors}{Zirui CHEN et al.}

\begin{abstract}
  Third-party libraries are essential in software development as they prevent the need for developers to recreate existing functionalities. However, vulnerabilities within these libraries pose significant risks to dependent projects. \xx{Upgrading dependencies is infeasible when no patched version is available or when downstream projects have strict version constraints.} Moreover, repairing the vulnerability proves challenging when the source code of the library is inaccessible. State-of-the-art automatic vulnerability repair and automatic program repair methods both fail to address this issue. Therefore, mitigating library vulnerabilities \xx{without access to source code or available patches} is crucial for a swift response to potential security attacks. Existing tools encounter challenges concerning generalizability and functional security. \dx{Upon vulnerability disclosure, \appname prepares a pool of candidate mitigation strategies by retrieving explicit workarounds from public sources and supplementing the pool with generated strategies based on vulnerability reproduction behavior. \appname then selects a suitable strategy, autonomously gathers repository-specific context, and adaptively generates a validated mitigation patch.} \xx{We evaluate \appname on 474 impacted functions associated with 94 vulnerabilities. \appname successfully mitigates 393 functions, achieving a success rate of 82.91\%,} which substantially outperforms the adapted baseline in blocking vulnerability reproduction under our exploit-based evaluation. Additionally, our ablation study evaluates the contributions of strategy composition, strategy selection, and repository-level context collection. Further evaluations show that \appname{} can mitigate vulnerabilities
disclosed after the knowledge cutoff of the base model.
\end{abstract}




\maketitle

\section{Introduction}

% 开源依赖漏洞的问题.pre-patch window, updating
Third-party libraries play a crucial role in modern software development ~\cite{kula2018developers, Markus1, Synopsys1,chen2026empirical}, as evidenced by their incorporation in 96\% of software projects~\cite{Synopsys1}. They significantly boost software development efficiency by facilitating the reuse of standard functionalities~\cite{Wang1}, thereby economizing on time and resources~\cite{kula2018developers, Yuan1}. Despite their benefits, third-party libraries are inherently exposed to design flaws, a prevalent issue in software ecosystems~\cite{Bavota1, kula2018developers, pashchenko2018vulnerable}. Projects dependent on vulnerable libraries are exposed to potential risks stemming from vulnerability propagation ~\cite{Decan1, Mir1}.  For instance, in 2021, the National Vulnerability Database (NVD) revealed a Remote Code Execution (RCE) vulnerability in Apache Log4j2 ~\cite{report4j}, impacting millions of projects and necessitating prompt action from project maintainers ~\cite{zirui2024exploiting}. What makes the situation worse is that a common practice in companies is to release only the binary of a library, which complicates library vulnerability detection in the supply chain scenario due to the lack of source code~\cite{Zhan2024Patch, Yang2023Day}.

% repair不行
% 我们尝试修复并且阻止漏洞的调用

% 这里改为 我们调研了sota的avr工作，但是发现其在直接修复**不行
Automatic program repair (APR) and automatic vulnerability repair (AVR) have been extensively studied to address vulnerabilities~\cite{Gao2019Repair, Michael2022Repair, Gao2021Repair, Hong2020Repair, Huang2019Repair, Michael2024Repair, Pearce2023Repair, Zhou2024Vulmaster}. However, to the best of our knowledge, no existing method can patch a binary library file based solely on the vulnerability description. In downstream projects,  developers interact with dependent libraries using API calls. When the source code of the library is unavailable, the lack of sufficient vulnerability context makes it difficult to address security issues in downstream projects. We further examine why APR and AVR methods, such as \master ~\cite{Zhou2024Vulmaster} and \think ~\cite{Yin2024ThinkRepair}, fail to address library vulnerabilities.
% 在我们的experiment和discussion章节我们证明了现有的AVR工作，如VulMaster在本任务上的表现不尽人意。
To mitigate this issue, \textbf{we aim to mitigate upstream vulnerabilities in downstream projects before official patches, even in cases where only upstream binary files are available}.

% implement thread watchdogs to cap and timeout parse runtimes.
% https://nvd.nist.gov/vuln/detail/CVE-2021-37714

% 解决依赖库漏洞不能仅依靠升级版本
% 对版本有特定的要求
% todo: 讲解pre-patch windows
% 1. patch之前，那么确实没有办法更新（这是安全那篇文章提出的概念）
% 2. patch之后，更新依赖版本会导致breaking change（这是软工主流关注的问题）
% 那么不管在哪种情况下，基于依赖fix的修复都不是靠谱的，（1里面fix缺失，2里面fix有但是打不进去）

When a library vulnerability is disclosed, project maintainers should neutralize its impact~\cite{Iannone1}. Existing studies propose various methods to address library vulnerabilities, such as updating library versions~\cite{Soto1, Zhang1}, migrating security patches~\cite{ mitigation, pan2024automating},  and generating workarounds~\cite{huang2019rapid, Huang2016Neu, zeng2018codelesspatchingheapvulnerabilities}.
Updating the vulnerable dependency to a patched version stands out as a widely accepted strategy for preventing the exploitation of the vulnerability~\cite{Soto1, Zhang1}. However, neutralizing vulnerabilities presents substantial challenges for projects with specific version requirements~\cite{Wang1, kim1, Bogart2}. For instance, \textit{Security Onion}, a log management platform, faced difficulties in resolving CVE-2021-44228 due to the utilization of outdated Log4j2 versions within its standard \dx{Docker} images~\cite{OrionDiscussion}. Recent studies  have centered on strategies without updating the version to neutralize vulnerabilities, such as \xx{backporting security patches from newer versions to older stable versions} ~\cite{mitigation, pan2024automating}, negating the necessity to update the dependency version~\cite{ mitigation}. However, these tools mainly rely on addressing the vulnerability only after it has been fixed, which leaves systems open to attacks before remediation ~\cite{Huang2016Neu, huang2019rapid}. Several studies have proposed tools to mitigate vulnerabilities when patches are unavailable~\cite{ huang2019rapid, Huang2016Neu, zeng2018codelesspatchingheapvulnerabilities}. These tools shield the vulnerability from attackers by blocking the execution of vulnerable code~\cite{huang2019rapid}.  

Existing methods face two major challenges: \ding{182} how to address the impact of vulnerabilities without existing fixes or source code of libraries and  \ding{183} how to generate mitigations \xx{without affecting existing functionality}. Therefore, we require a method to universally generate mitigations without impacting existing functionalities. We find that libraries propose certain vulnerability mitigation strategies, such as security configurations~\cite{mitigate44228}. Additionally, general strategies can be applied to mitigate specific types of vulnerabilities~\cite{CVE-2021-37714, Huang2016Neu, CVE-2022-40151, CVE-2024-23684}. These strategies effectively mitigate library vulnerabilities while avoiding the challenges faced by existing methods. The code generation capabilities of large language models (LLMs) enable the integration of mitigation strategies with impacted functions from various projects.  Therefore, we explore leveraging LLMs and existing strategies to neutralize library vulnerabilities.

% 我们发现，依赖库本身会提出一些漏洞的缓解策略，如安全设置~\cite{Mitigate44228}，同时在特定类型的漏洞可以采用通用的策略解决~\cite{CVE-2021-37714, Huang2016Neu, CVE-2022-40151, CVE-2024-23684}。上述策略可以有效解决现有方法面临的挑战，而大语言模型可以将上述策略与下游项目待缓解的代码结合，生成安全的代码，因此我们探寻基于大语言模型与现有策略进行功能安全的依赖库漏洞缓解。

\dx{We propose an LLM-based method, \appname (\textbf{L}ibrary v\textbf{U}lnerability \textbf{M}itigation g\textbf{EN}eration), designed to neutralize library vulnerabilities in downstream projects without modifying the library source code. \appname operates through an iterative process comprising four core modules: Mitigation Strategy Preparation, Mitigation Strategy Selection, Mitigation Context Collection, and Adaptive Mitigation Generation. First, it prepares a candidate strategy pool by aggregating explicit public mitigation knowledge and generating strategies based on vulnerability reproduction behavior. Next, \appname chooses the most suitable strategy based on existing context and historical effectiveness, followed by a tool-augmented repo-level context collection phase. Finally, \appname generates structured mitigation patches and validates them against both downstream compatibility and security effectiveness, utilizing the verification feedback to incrementally refine the strategy selection and patch generation in subsequent attempts.}

% 方法表现
We assess the effectiveness of \appname in addressing vulnerabilities disclosed between 2014 and 2024. Our dataset comprises \dx{474} impacted functions extracted from projects reliant on \dx{48} vulnerable \dx{Java} libraries, covering \dx{94} vulnerabilities and \dx{35} CWEs. \appname effectively neutralizes \dx{82.91\%} of these functions, surpassing \baselinename, the state-of-the-art method for security workaround generation, which achieves \xx{34.60\%}. We conduct an ablation study to demonstrate the effectiveness of the proposed strategy. To assess the impact of \appname on the original functionality, we compile a dataset comprising \xx{6,377} tests for the impacted functions. \xx{The tests comprise 473 constructed tests, 3,943 EvoSuite-generated tests, and 1,961 original project tests.} After mitigation, 6,272 tests remained consistent, corresponding to a consistency rate of 98.35\%. Compared to the base model, \appname demonstrates better performance, validating the effectiveness of its design.

% We believe that mitigating vulnerabilities from upstream dependency libraries in downstream projects is a valuable research direction.

% 我们工作的贡献
% 1. 我们的方法在项目级可以实现快速止损，在漏洞修复前。
% 2. 我们提供了一个数据集
The main contributions of this paper are summarized as follows: 
\begin{itemize}[leftmargin=*]

\item {We propose an LLM-based approach, named \appname, designed to address library vulnerabilities \xx{without requiring available patches or access to the source code of vulnerable libraries}.}

\item {We construct a dataset comprising \dx{94} Java CVE vulnerabilities across \dx{35} CWEs, encompassing \dx{474 impacted functions with corresponding exploit tests} and \xx{6,377} functionality tests.}

\item {\dx{\appname effectively neutralized 393 (82.91\%) vulnerable functions in downstream projects while maintaining a \xx{98.35}\% pass rate in functional testing, highlighting its robust capability with limited compatibility impact.}}

\end{itemize}

% 我们文章的架构
% This paper is structured as follows. In Section 2, we present a preliminary study that illustrates vulnerability reproducing behaviors and the mitigation of library vulnerabilities. In Section 3, we describe the implementation of our methodology. In Section 4, we demonstrate the effectiveness of our approach through experimental evaluation. In Section 5, we discuss the threat to validity of our proposed method. In Section 6, we offer a discussion of the related work. Finally, in Section 7, we summarize our method and mention our future works.

\section{Preliminary Study}

In this section, we first conduct an empirical study on the types of library vulnerabilities and propose a classification criterion. Then, we present two motivating examples.

\subsection{Vulnerability Reproducing Behavior}

Previous research on vulnerabilities in dependency libraries has summarized the effects of these vulnerabilities~\cite{zirui2024exploiting, Hong2022Mimicry}, including reproducing behaviors such as `wrong functional behavior' and `out of memory'. However, they have not discussed the generalizability of the behavior-based vulnerability classification criteria. We conduct a large-scale empirical study to gain a more comprehensive understanding of library vulnerability-reproduction behaviors. To systematically gather historical vulnerabilities, we collect the top 10 most-used dependencies and versions with the highest usage from each top category of Java libraries in the Maven Central Repository. In total, we identify 163 vulnerabilities with the vulnerable versions accounting for over 160,000 usages.

\begin{table}[tb]
    \centering
    \caption{Library Vulnerability Reproducing Behaviors.}
    \label{tab: behaviors}
    \footnotesize
    \setlength{\tabcolsep}{6pt}
    \renewcommand{\arraystretch}{1.0}
    \begin{tabular}{>{\raggedright\arraybackslash}p{3.6cm} c >{\raggedright\arraybackslash}p{3.6cm} c}
        \toprule
        \textbf{Behaviors (Prior Study)} & \textbf{Vul.} & \textbf{Behaviors (Uncovered)} & \textbf{Vul.} \\
        \midrule
        \rowcolor[gray]{0.95} Wrong Functional Behavior & 18 & Information Leakage & 6 \\
        \rowcolor[gray]{0.9} Out of Memory & 10 & Crash & 3 \\
        \rowcolor[gray]{0.9} Exception & 8 & Stack Overflow & 5 \\
        \rowcolor[gray]{0.85} Remote Code Execution & 59 & Cross-Site Scripting & 8 \\
        \rowcolor[gray]{0.85} XXE Injection & 5 & Path Traversal & 5 \\
        \rowcolor[gray]{0.85} Wrong File Permissions & 5 & Access Internal Resource & 3 \\
        \rowcolor[gray]{0.85} SQL Injection & 2 & XML Injection & 1 \\
        \rowcolor[gray]{0.78} Infinite Loop & 9 & CPU Consumption & 9 \\
        \multicolumn{2}{c}{} & Other Behaviors & 7 \\
        \midrule
         Within Prior Categories & 116 &  Outside Prior Categories & 47\\ 
        \bottomrule
    \end{tabular}

    \vspace{1mm}
    {\footnotesize\textit{The gray areas represent vulnerabilities covered by our types.}}
\end{table}

We classify the collected vulnerabilities based on the behaviors outlined by Kang et al. ~\cite{Hong2022Mimicry}, as shown in Table \ref{tab: behaviors}. The table's left side represents the vulnerabilities that existing categories can successfully cover.  The classification method proposed in the prior research successfully covers 116 out of 163 vulnerabilities, which preliminarily supports the rationale of classifying vulnerabilities based on reproducing behaviors. However, existing categories are not comprehensive enough, resulting in a low coverage rate. For example, in Java, Stack Overflow, Out of Memory, Crash, and Exception can all be considered as types of error/exception throwing; Infinite Loop and CPU Consumption can be viewed as the exhaustion of specific resources.

Based on these findings, we categorize library vulnerabilities into four types: Uncaught Exception, Resource Exhaustion, Malicious Code Execution, and Wrong Return Value. This classification criterion successfully covers 156 out of 163 vulnerabilities in widely used libraries: Uncaught Exception (26), Resource Exhaustion (18), Malicious Code Execution (88), and Wrong Return Value (24). In Table \ref{tab: behaviors}, we use different shades of gray to represent the vulnerabilities covered by the types we proposed, ranging from light to dark: Wrong Return Value, Uncaught Exception, Malicious Code Execution, and Resource Exhaustion. However, this classification method does not adequately address less common attack types, such as DNS Poisoning and Timing Attacks. \dx{Therefore, our study needs a strategy generation process that can support mitigation beyond predefined types.}

\subsection{Motivating Examples}

\dx{Downstream projects often need to neutralize vulnerable dependencies before an official fix is available or before upgrading the library is feasible. However, practical downstream mitigation faces two key challenges. First, even when an existing mitigation strategy is publicly available, developers still need to determine how to integrate it into the downstream project, which requires locating the appropriate mitigation points in the downstream code and configuration. Second, many vulnerabilities do not provide any explicit mitigation strategy at all, making it impossible to rely solely on existing workaround knowledge. These challenges suggest that effective downstream mitigation requires both the ability to \ding{182} \textbf{adapt existing strategies to the project context} and the ability to \ding{183} \textbf{infer mitigation strategies when no prior workaround is available}.}

\begin{figure}[tb]
  \centering
  \begin{minipage}[t]{0.48\linewidth}
    \centering
    \includegraphics[width=\linewidth]{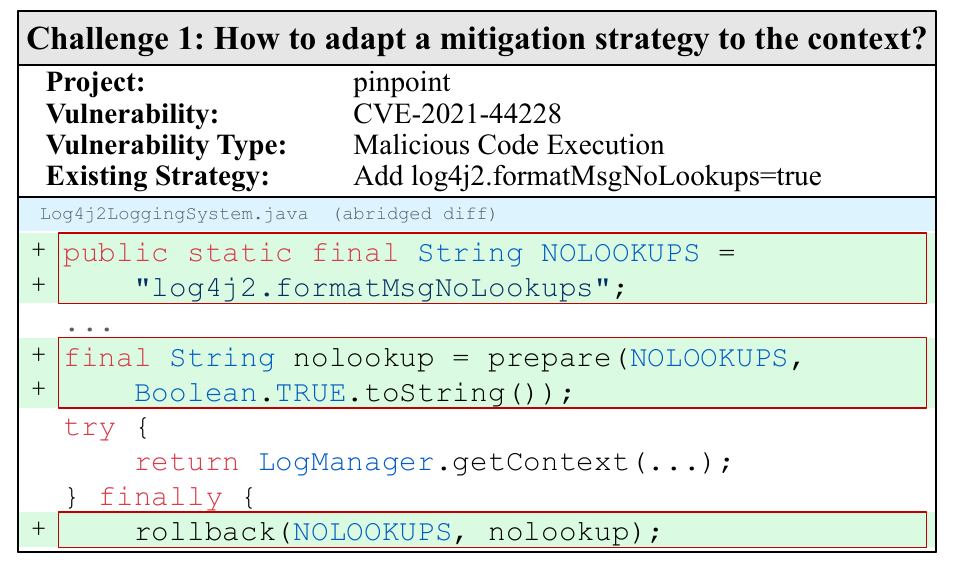}
    \footnotesize \textbf{(a)} Adapting an Existing Workaround
  \end{minipage}
  \hfill
  \begin{minipage}[t]{0.48\linewidth}
    \centering
    \includegraphics[width=\linewidth]{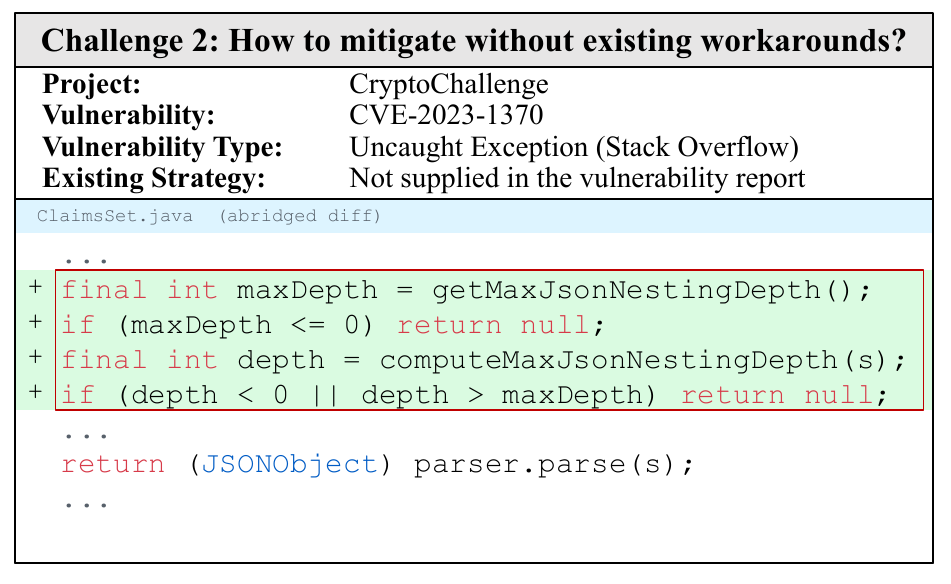}
    \footnotesize \textbf{(b)} Generating a Suitable Mitigation Strategy
  \end{minipage}
  \caption{Two motivating examples of downstream vulnerability mitigation (Adapting an existing workaround to the repository context and inferring a mitigation strategy rather than employing existing workarounds).}
  \Description{Two panels show project information and abridged code
diffs. Panel (a) shows the Pinpoint workaround for CVE-2021-44228,
setting a Log4j property before logger initialization and restoring
it afterward. Panel (b) shows the CryptoChallenge mitigation for
CVE-2023-1370, checking JSON nesting depth before parsing.
Green shading and red outlines highlight the added mitigation logic.}
  \label{fig: Mitigation}
\end{figure}

\begin{itemize}[leftmargin=1.2em]
    \item \dx{Existing mitigation knowledge is often available only as a high-level workaround, while its actual application still depends on the downstream project context. Consider CVE-2021-44228 as an example. NVD provides mitigation references indicating that downstream projects can prevent exploitation by enabling the \textit{log4j2.formatMsgNoLookups} property~\cite{mitigate44228}. However, this guidance does not directly specify where the mitigation should be inserted in a concrete project. As shown in Figure~\ref{fig: Mitigation}(a), in the downstream project \textit{pinpoint}, the mitigation is realized by introducing the configuration string \texttt{String NOLOOKUPS = ``log4j2.formatMsgNoLookups''} at the appropriate project location, so that the vulnerable behavior can be neutralized without immediately upgrading the \textit{Log4j} dependency. This example highlights that, even when an existing workaround is known, effective downstream mitigation still requires identifying suitable mitigation points and adapting the strategy to project-specific code or configuration. To support this adaptation process, \appname performs repository-level context collection.}

    \item A vulnerability advisory may only explain how exploitation occurs without providing a downstream mitigation recipe. Figure~\ref{fig:motivation}(b) shows a downstream mitigation for CVE-2023-1370 in \textit{CryptoChallenge}, which uses \textit{json-smart}. Deeply nested JSON inputs can cause stack exhaustion during
recursive parsing.
The mitigation computes the maximum nesting depth of the input
in \texttt{ClaimsSet.java} and rejects inputs exceeding the
configured limit before invoking \texttt{JSONParser.parse()}.
This example motivates combining general mitigation principles
with repository-specific context to temporarily block exploitation
when the advisory provides no explicit downstream workaround.
\end{itemize}

These examples motivate two complementary capabilities:
adapting existing workarounds to repository-specific
mitigation points and inferring mitigation strategies
from vulnerability types and manifestations when no workaround
is available.

\section{Proposed Approach}

\begin{figure*}[htbp] 
  \centering	
  \resizebox{\linewidth}{!}{
  \includegraphics[width=1\linewidth]{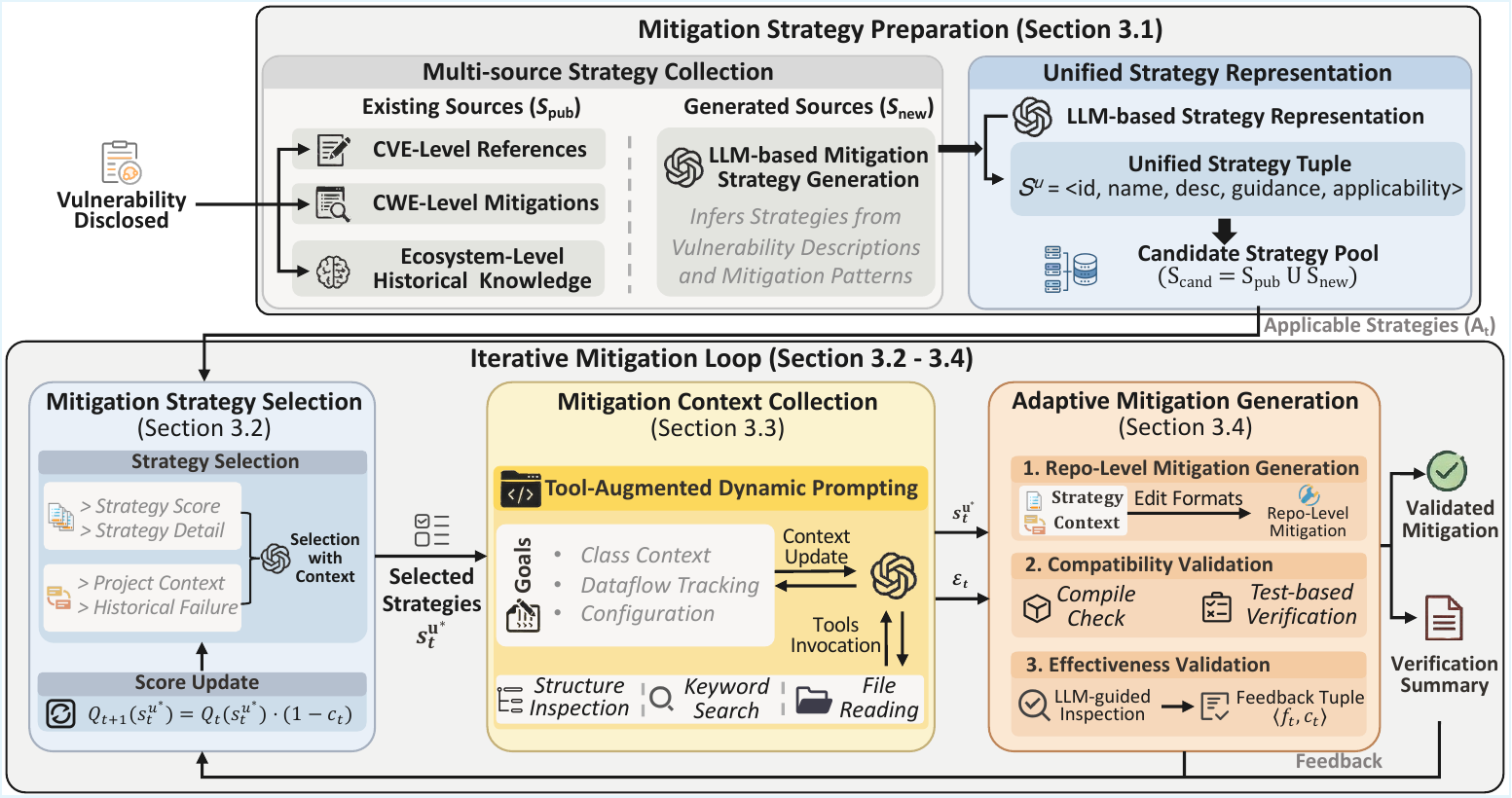}
  }
  \caption{Overall framework of \appname.}
  \label{fig: overview}
  \Description{Workflow diagram of LUMEN, consisting of strategy
preparation and an iterative mitigation loop. Strategy preparation
combines existing mitigation knowledge and LLM-generated strategies
into a unified candidate pool. The loop selects applicable strategies,
collects repository context through tool-augmented prompting, and
generates repository-level mitigations. Compatibility and effectiveness
validation provide feedback to update strategy scores and guide
subsequent iterations. The outputs are a validated mitigation and
a verification summary.}
\end{figure*}

In this section, we present our proposed approach, \textit{\appname}. As shown in Figure \ref{fig: overview}, \dx{\appname comprises four core modules: \textbf{Mitigation Strategy Preparation}, \textbf{Mitigation Strategy Selection}, \textbf{Mitigation Context Collection}, and \textbf{Adaptive Mitigation Generation}. To mitigate library vulnerabilities in real-world downstream projects, \appname first prepares a pool of candidate mitigation strategies from multiple sources (Section~\ref{3.1}). It then performs mitigation through an iterative loop of strategy selection (Section~\ref{3.2}), context collection (Section~\ref{3.3}), and mitigation generation (Section~\ref{3.4}), where each generated mitigation is validated against both effectiveness and functionality preservation.}

\subsection{Mitigation Strategy Preparation}
\label{3.1}
\dx{After a vulnerability is disclosed, \appname first collects publicly available mitigation knowledge from multiple sources, transforms it into a unified strategy representation, and refines the resulting candidates according to their applicability to downstream projects. To address cases where existing mitigation knowledge is unavailable or inapplicable to the downstream conditions, \appname further supplements the strategy pool with inferred strategies derived from the vulnerability description.}

\subsubsection{Multi-source Strategy Collection}

\dx{To ensure comprehensive coverage of mitigation knowledge, \appname constructs a candidate strategy pool, denoted as $S_{cand}$, by collecting evidence across explicit public sources and implicit generative sources. Formally, $S_{cand} = S_{pub} \cup S_{new}$. For the publicly available sources ($S_{pub}$), \appname extracts knowledge across three dimensions: $S_{pub} = S_{cve} \cup S_{cwe} \cup S_{eco}$. \ding{182} First, it populates $S_{cve}$ by extracting \textbf{CVE-Level} evidence directly from the target vulnerability report, capturing explicit workarounds detailed in the description and reference links. \ding{183} Second, it populates $S_{cwe}$ by retrieving \textbf{CWE-Level} mitigation practices formally documented within the \textit{Potential Mitigations} section of the corresponding CWE report. In total, 673 CWEs contain such mitigation information. If the target CWE does not provide this section, \appname falls back to its parent CWE under the CWE-1000 Research Concepts view to obtain a higher-level mitigation strategy. \ding{184} Third, $S_{eco}$ captures \textbf{Ecosystem-Level} historical knowledge by reusing mitigations from prior vulnerabilities within the same library, recognizing that libraries often manage risky functionalities through specific security configurations.}

\dx{However, publicly disclosed mitigation knowledge is not always available or applicable. In scenarios where explicit links are absent, or CWE-level guidance is too abstract to support downstream mitigation (i.e., $S_{pub}$ yields insufficient viable strategies), \appname initiates a strategy generation phase to populate $S_{new}$. In this phase, the LLM infers candidate mitigation strategies directly from the vulnerability description and its observed behavior.}

\dx{To enhance the reliability of mitigation generation, we summarize a set of mitigation patterns based on our preliminary study. \appname first evaluates whether the vulnerability behavior matches one of these predefined patterns. \ding{182} If a match is found and the corresponding strategy is deemed applicable to the current downstream context, \appname adds it to $S_{new}$. \ding{183} Otherwise, if no predefined pattern applies or the suggested mitigation is judged ineffective, \appname falls back to generating $S_{new}$ prompted solely by the vulnerability description. We define the following mitigation \xx{patterns}:}

\begin{itemize}[leftmargin=*]

\item \textbf{Exception Catching.} A vulnerability leading to an \textit{Uncaught Exception} means the library unexpectedly throws an error while manifesting the vulnerability, causing the dependent project to fail and resulting in a denial-of-service condition. In this scenario, we adopt the recommended workaround from CVE-2022-40151~\cite{CVE-2022-40151}, a vulnerability in XStream that can lead to a DoS attack through a stack overflow. We propose synthesizing a block to catch the unexpected exception within the client code that invokes the vulnerable function. 

\item \textbf{Thread Monitoring.} \textit{Resource Exhaustion} refers to the depletion of resources such as time, CPU, and memory due to vulnerabilities that lead to infinite function loops or excessive CPU usage. To address this issue, we introduce a thread watchdog mechanism to monitor resource consumption during runtime, a mitigation approach demonstrated in handling CVE-2021-37714~\cite{CVE-2021-37714}.  The watchdog requests interruption when execution exceeds the
configured time bound. This strategy can terminate excessive
computation when the underlying task responds to interruption.

\item \textbf{Input Validation.} This pattern primarily addresses vulnerabilities manifesting as \textit{Malicious Code Execution} or \textit{Wrong Return Value}. \textit{Malicious Code Execution} encompasses attacks such as arbitrary code injection, SQL injection, and XML External Entity (XXE) Injection. The specific behavior heavily relies on the input provided. We focus on extracting critical details regarding input characteristics that trigger the exploit. For instance, in CVE-2021-44228, the attack is \xx{triggered} by loading code from LDAP servers; thus, inputs containing substrings like ``\texttt{ldap://}'' possess high exploit potential. Following the approach outlined in CVE-2024-23684~\cite{CVE-2024-23684}, we deploy an input validator to identify and intercept potentially compromised inputs before they reach the vulnerable library function.

\end{itemize}

\subsubsection{Unified Strategy Representation}

\dx{The candidate strategy pool $S_{cand}$ constructed in the previous stage aggregates mitigation knowledge from diverse sources. However, these strategies are not directly usable for downstream mitigation generation, as they are often unstructured, redundant, and inapplicable to the downstream project.}

\dx{To address these issues, \appname transforms each candidate strategy into a unified structured representation and filters out strategies that are not applicable to the downstream setting. Specifically, given a candidate strategy $s \in S_{cand}$, \appname normalizes it into the following tuple:
\[
s^{u} = \langle id, name, desc, guidance, applicable \rangle,
\]
where $id$ is the unique identifier of the strategy, $name$ is a concise strategy name, $desc$ is a natural-language description of the strategy intent, $guidance$ provides step-by-step implementation guidance for realizing the mitigation, and $applicable \in \{\texttt{true}, \texttt{false}\}$ indicates whether the strategy is suitable for the downstream project context.}

\begin{mdframed}[
  linecolor=myblue!50,
  linewidth=2pt,
  roundcorner=10pt,
  backgroundcolor=myyellow!20,
  suppressfirstparskip=false
]
\small
\hspace{-0.5mm}\textbf{Prompt for Mitigation Strategy Pool Generation:}  

\hspace{-4mm}You are generating a strategy pool for mitigating a library vulnerability on downstream projects. 

\hspace{-4mm}\textbf{Input:} \textit{[Vulnerability Description]}, \textit{[Existing Strategies]}, and \textit{[Generated Strategies]}.

\hspace{-4mm}\textbf{Task:} Produce candidate mitigation strategies applicable to the downstream project WITHOUT upgrading the vulnerable dependency. Only include strategies implementable by editing the downstream repository.

\hspace{-4mm}\textbf{Output Requirements:} Formalize each strategy as a unified tuple $s^{u}$.

\hspace{-4mm}\textbf{[Definitions of each component in the tuple]}
\end{mdframed}

% 这里，就是我们用LLM，1. 选择了下游的策略。2。把策略加工成了指定的格式

\subsection{Mitigation Strategy Selection}
\label{3.2}

\subsubsection{\xx{Feedback-Guided Strategy Selection.}}

Following the unified representation, \appname first filters out strategies that are not applicable to the downstream project and retains only viable candidates. At mitigation attempt $t$, we denote the available strategy set as
$A_t = \{s^u \mid s \in S_{cand} \wedge s^u.applicable = true\}$.
The goal of this step is to select a strategy that is both compatible with the current downstream context and unlikely to repeat previously observed failures.
\xx{For each candidate strategy $s^u \in A_t$, \appname maintains a preference score $Q_t(s^u)$ and a failure memory $H_t(s^u)$. The score $Q_t(s^u)$ summarizes how reliable the strategy has appeared during the current mitigation process, while $H_t(s^u)$ records textual feedback from previous failed attempts using this strategy. Initially, all applicable strategies are assigned the same score,
i.e., $Q_1(s^u)=1.0$, and their failure memories are empty.}

\xx{After each failed attempt, \appname updates the preference score
and failure memory of the selected strategy. These accumulated signals are then provided to the LLM to guide subsequent selections, allowing \appname to avoid repeatedly applying strategies that have already caused high-confidence failures while still considering strategies that may remain suitable under a different patch instantiation. Specifically, \appname provides the LLM with the downstream context $x_t$, the candidate strategies in $A_t$, their current scores $Q_t(s^u)$, and their previous failure feedback $H_t(s^u)$. The LLM then selects the strategy whose mitigation guidance best matches the current vulnerability and repository context while avoiding strategies that have already led to high-confidence failures. Formally, we denote this selection process as:
\[
\mathcal{M}_t = \{(s^u, Q_t(s^u), H_t(s^u)) \mid s^u \in A_t\},
\quad
s_{t}^{u*} = \phi_{\mathrm{LLM}}(x_t, \mathcal{M}_t),
\]
where $\mathcal{M}_t$ denotes the candidate-strategy memory at attempt $t$, including each candidate strategy, its current preference score, and its previous failure feedback. $\phi_{\mathrm{LLM}}$ denotes the LLM-based context-aware selection function, which selects one strategy by jointly considering the downstream context, the accumulated preference score, and the failure history of each candidate strategy. In this way, \appname can prioritize strategies that better match the current vulnerability context while avoiding strategies that have already caused failures in previous attempts.}

\subsubsection{\xx{Strategy Feedback Update.}}

\xx{After executing the selected strategy $s_{t}^{u*}$, \appname first checks whether the generated patch compiles, and then uses LLM-guided inspection to assess functionality preservation and security effectiveness. If the patch passes all validation checks, \appname terminates and returns the validated mitigation.} Otherwise, the verification module returns a feedback tuple $\langle f_t, c_t \rangle$, where $f_t$ is a textual explanation of the failure and $c_t \in [0,1]$ is a confidence score estimating how likely the failure should be attributed to $s_{t}^{u*}$ being unsuitable for the current context. Based on this confidence score, \appname decreases the historical score of the selected strategy as follows:
\[
Q_{t+1}(s_t^{u*}) = Q_t(s_t^{u*}) \cdot (1 - c_t).
\]

In addition to updating the score, \appname appends the textual failure rationale $f_t$ to the failure memory of the selected strategy: \xx{
$H_{t+1}(s_{t}^{u*}) = H_t(s_{t}^{u*}) \cup \{f_t\}$}.
The updated score and failure memory are then provided to the LLM in subsequent strategy-selection and mitigation-generation attempts. \xx{This feedback mechanism helps \appname avoid repeatedly applying ineffective strategies while still allowing a strategy to be reconsidered when previous failures are likely caused by implementation-level issues rather than strategy-level mismatch.}

\subsection{Mitigation Context Collection}
\label{3.3}
After selecting a mitigation strategy, \appname{} collects the
repository-specific context required to instantiate it.
The prompt defines \appname{} as a downstream vulnerability
mitigation analyst. Given basic project information, including
the location and abstract syntax tree (AST) of the function
invoking the vulnerable library API, \appname{} uses
tool-augmented dynamic prompting to identify and gather
missing evidence.

\subsubsection{Goals}

\dx{We define three goals for the context collection stage. Together, these goals provide the context required to apply a selected mitigation strategy in the downstream repository.}

\begin{itemize}[leftmargin=*]
  \item \textbf{\dx{Object Context.}} 
    \dx{\appname traces the lifecycle of the target vulnerable object before the vulnerable API invocation. Specifically, it focuses on where the vulnerable class is instantiated and extracts subsequent configurations applied to it. \xx{This information is critical for determining appropriate mitigations.}}

    \item \textbf{\dx{Data-flow Tracking.}}
    \dx{\appname traces how attacker-controlled inputs flow to the vulnerable API call site. Rather than following a fixed-depth analysis, it explores the relevant call chain on demand to understand how external inputs reach the vulnerable operation. The exploration stops once sufficient context has been collected to identify an appropriate location for inserting mitigation logic.}

    \item \textbf{\dx{Configuration Context.}}
    \dx{\appname collects library configuration files and environment settings that may affect mitigation across the project. This includes build files (e.g., \textit{pom.xml}), framework configuration files, and other settings that define the runtime behavior of the library (e.g., \textit{antisamy.xml} for AntiSamy). Such information is essential for generating effective mitigations, especially in cases where mitigation must be realized by modifying the configuration rather than the vulnerable API usage.}
\end{itemize}

\subsubsection{Available Tools}
\dx{To support repository-level context collection, \appname includes the following tools for repository inspection.}

\begin{itemize}[leftmargin=*]
    \item \textbf{\dx{Project structure inspection.}}
    \dx{This tool allows \appname to inspect the directory structure of the downstream project at the depth required for the current analysis.}

    \item \textbf{\dx{Keyword search.}}
    \dx{This tool allows \appname to search the repository for mitigation-relevant identifiers, such as vulnerable API names, class names, or configuration options.}

    \item \textbf{\dx{File reading.}}
    \dx{This tool allows \appname to read arbitrary portions of a file, or the entire file when needed, to obtain mitigation-relevant information. }
\end{itemize}

\subsubsection{State Description}

\dx{During context collection, \appname maintains an explicit state to guide repository exploration. This state summarizes what information is required by the selected mitigation strategy, what mitigation-relevant evidence has already been collected, and whether the current context is sufficient for mitigation generation.}

\dx{Formally, let $\mathcal{G}=\{g_1,\dots,g_m\}$ denote the predefined collection goals, where each goal specifies a category of mitigation-relevant context required for downstream mitigation generation. At interaction step $t$, let $\mathcal{E}_t$ denote the set of repository evidence collected so far, and let $\mathcal{Q}_t \subseteq \mathcal{G}$ denote the queue of currently unsatisfied goals. \appname first selects a goal $g_t \in \mathcal{Q}_t$ for the current round of exploration, and then performs the following process.}

\begin{itemize}[leftmargin=*]
    \item \textbf{\dx{Generate collection tasks.}}
    \dx{In this state, \appname derives a concrete collection task $\tau_t$ based on the selected goal $g_t$ and the current evidence set $\mathcal{E}_t$. Unlike the predefined goals, which specify high-level categories of required context, $\tau_t$ represents the next immediate exploration task for the current round. }

    \item \textbf{\dx{Collect required context.}}
    \dx{To fulfill the current task $\tau_t$, \appname autonomously selects and invokes repository-inspection tools to gather additional evidence. The collected observations are incorporated into the evidence set. This process continues until $\tau_t$ is considered resolved or the budget for the current round is exhausted.}

    \item \textbf{\dx{Assess goal satisfaction.}}
    \dx{In this state, \appname evaluates whether the selected goal $g_t$ has been satisfied by the updated evidence set. If $g_t$ is not satisfied, \appname starts another round by generating a new collection task for the same goal. If $g_t$ remains unsatisfied after a predefined number of rounds, it is returned to the unsatisfied-goal queue for future exploration. }
\end{itemize}

\dx{The context-collection process terminates when the unsatisfied-goal queue $\mathcal{Q}_t$ becomes empty, or the predefined interaction budget is reached. In either case, the collected evidence is retained as the final context for the current attempt. This design bounds the exploration cost while still allowing \appname to generate mitigations based on the most relevant repository-level information gathered so far.}

\subsection{Adaptive Mitigation Generation}
\label{3.4}

\dx{The initial patch generated by the LLM may still require adaptation before it becomes a valid downstream mitigation. Therefore, \appname performs an iterative adaptation process to refine the generated patch under two complementary objectives: preserving downstream compatibility and ensuring security effectiveness. These two objectives are validated separately, and the resulting feedback is used to guide subsequent patch refinement.}

\subsubsection{Mitigation Generation}

\dx{Given the selected unified strategy $s_t^{u*}$ and the repository-specific evidence collected in Section~\ref{3.3}, \appname next generates a downstream mitigation patch tailored to the target project. To achieve this, \appname constructs a generation prompt by combining three categories of information: \ding{182} the vulnerability description, \ding{183} the selected strategy $s_t^{u*}$, including its description and actionable guidance, and \ding{184} the collected project-specific context. Based on this combined input, the LLM produces a candidate mitigation patch $p_t$ for the current attempt.}

\dx{LLMs often struggle with numerical tasks and may produce incorrect numerical specifications~\cite{yu2025PATCHAGENT}. As a result, \appname does not directly rely on diff-style patch generation, since such formats typically require precise numeric information, such as line offsets and edit locations, which LLMs are prone to generating inaccurately, thereby reducing patch reliability. Instead, following prior work~\cite{Bouzenia2025Repairagent}, \appname constrains the generation output using a predefined structured edit format. Under this format, the LLM is required to express each candidate patch as a sequence of explicit repository edits, where each edit specifies the target file, the edit action, the exact target snippet when needed, and the replacement or inserted content. This format supports common modification operations, including file creation, exact replacement, deletion, and insertion relative to an anchor snippet rather than line numbers. By enforcing this structured patch format, \appname can localize intended modifications more precisely and reduce ambiguity during patch application.}

\begin{mdframed}[
  linecolor=myblue!50,
  linewidth=2pt,
  roundcorner=10pt,
  backgroundcolor=myyellow!20,
  suppressfirstparskip=false
]
\small
\hspace{-0.5mm}\textbf{Prompt for Adaptive Mitigation Generation:}  

\hspace{-4mm}You are generating precise code patches to mitigate a library vulnerability within a downstream project.

\hspace{-4mm}\textbf{Input Context}:
\begin{itemize}[leftmargin=4mm, topsep=0pt, partopsep=0pt, parsep=0pt, itemsep=0pt]
\item Selected Strategy ($s_t^{u*}$): The specific mitigation guidance and prior critic reflections.
\item Vulnerability Info: \textit{[Vulnerability Description]} and \textit{[Vulnerable API Locations]}.
\item Downstream State: \textit{[Collected Context]} and \textit{[Available Code-Exploration Tools]}.
\item Iterative Feedback: \textit{[Current Round Index]}, \textit{[Previous Failures]}, and \textit{[Verification Feedback]}.
\end{itemize}

\hspace{-4mm}\textbf{Task}: Apply the strategy to the project without upgrading the vulnerable library according to the context. 

\hspace{-4mm}\textbf{Output Requirements}: Produce exactly one patch candidate. The patch must be represented as a sequence of the given edit operations. Each edit must specify the file path, the exact target snippet, and the new content.

\hspace{-4mm}\textbf{[Predefined Edit Formats]}
\end{mdframed}

\subsubsection{Downstream Compatibility Validation}

To serve as a practical downstream mitigation, a generated patch must be applicable to the target repository and preserve its original functionality. \appname therefore first validates the candidate patch $p_t$ from the perspective of downstream compatibility.

This validation includes two stages. First, \appname checks whether the generated patch can be successfully applied and compiled in the target project. If patch application fails or the modified project cannot be built, \appname performs \textbf{syntax-level adaptation}. In this phase, the patch application error, compilation error, or build log is fed back to the LLM together with the faulty patch, and the LLM is prompted to revise the patch while preserving the intent of the selected strategy. This process is repeated until the patch becomes applicable and syntactically valid, or a predefined retry limit is reached.

Second, once the patch is successfully applied and compiled, \appname validates whether it preserves the existing functionality of the downstream project. Specifically, it executes the downstream test suite before and after applying the mitigation and compares the results. If new test failures are introduced, the failed test information is incorporated into the feedback prompt, and the LLM is asked to revise the patch so that it eliminates the regression while still maintaining the intended mitigation logic. Through this process, \appname incrementally improves the compatibility of the generated mitigation with the downstream project.

% \begin{mdframed}
% [linecolor=myblue!50,linewidth=2pt,roundcorner=10pt,backgroundcolor=myyellow!20]
% \small
% \hspace{-0.5mm}\textbf{Prompt for Adapting:}  

% \hspace{-4mm}The mitigated function should avoid influencing test: \textit{[Failed Test]}

% \hspace{-4mm}The following code contains a syntax error \textit{[Error Log]}, please fix it. \textit{[Error Function]}
% \end{mdframed}

\subsubsection{Security Effectiveness Validation}

\label{security}

\dx{Beyond compatibility, a valid downstream mitigation must also prevent the vulnerability from being triggered in the target project. \appname therefore separately validates the generated patch from the perspective of security effectiveness.}

\dx{Specifically, after the candidate patch passes compatibility validation, \appname performs prompt-guided vulnerability-oriented verification to assess its security effectiveness. Given the vulnerability description and the generated patch, \appname constructs a verification prompt that instructs the LLM to determine how the vulnerable behavior should be checked in the current downstream setting. Based on this prompt, the system verifies whether the mitigation successfully blocks the exploit trigger or prevents the undesirable behavior.}

\dx{If the candidate patch fails to preserve downstream compatibility or security effectiveness, \appname summarizes the verification outcome into a structured feedback tuple $\langle f_t, c_t \rangle$, where $f_t$ records the textual rationale of the failure and $c_t$ estimates how likely the selected strategy $s_t^{u*}$ is unsuitable for the current downstream context. This feedback is then returned to the strategy-selection module described in Section~\ref{3.2} to update the selected strategy, helping \appname avoid repeatedly instantiating ineffective mitigation strategies.}

\begin{mdframed}[
  linecolor=myblue!50,
  linewidth=2pt,
  roundcorner=10pt,
  backgroundcolor=myyellow!20,
  suppressfirstparskip=false
]
\small
\hspace{-0.5mm}\textbf{Prompt for Static Audit:}  

\hspace{-4mm}You are performing a static audit on a proposed vulnerability mitigation.

\hspace{-4mm}\textbf{Input Context}:
\begin{itemize}[leftmargin=4mm, topsep=0pt, partopsep=0pt, parsep=0pt, itemsep=0pt]
\item Vulnerability Info: \textit{[CWE Info]} and \textit{[CVE Info]}.
\item Code Changes: \textit{[Prior Downstream Code]} and \textit{[Mitigated Downstream Code]}.
\end{itemize}

\hspace{-4mm}\textbf{Task}: Analyze whether the applied mitigation successfully neutralizes the malicious attacks.

\hspace{-4mm}\textbf{Output Requirements}: Formalize the evaluation as an audit tuple $v = \langle verdict, confidence, reason \rangle$.
\begin{itemize}[leftmargin=4mm, topsep=0pt, partopsep=0pt, parsep=0pt, itemsep=0pt]
\item \textbf{verdict}: Whether the mitigation is secure against exploitation.
\item \textbf{confidence}: A probability score $c \in [0, 1]$ reflecting the certainty of the verdict.
\item \textbf{reason}: A justification for the audit result.
\end{itemize}
\end{mdframed}

\section{Experiment Setup}

We conduct an experimental evaluation to assess the effectiveness and reliability of our method in neutralizing library vulnerabilities. We aim to answer the following four RQs:

\begin{itemize}[leftmargin=*]
\item {\textbf{RQ1: What is the effectiveness of \appnamebold in mitigating library vulnerabilities? }}

\item {\textbf{RQ2: How effective are the key designs of \appnamebold?}}

\item {\textbf{RQ3: Does \appnamebold mitigate vulnerabilities without impacting functionality?}}

\item {\dx{\textbf{RQ4: To what extent can \appnamebold generalize to mitigating unseen vulnerabilities?}}}

\end{itemize}

In this section, we first discuss our experimental subjects, including the constructed dataset, evaluation metrics, and the baseline. Then, we describe our experiment settings.

\subsection{Data Collection} 
During our data collection process, we gather vulnerabilities with various weakness enumerations and the functions impacted by these vulnerabilities to evaluate the performance of \appname.

\noindent \subsubsection{Vulnerabilities.} We collect vulnerabilities disclosed in Java libraries from 2014 to 2024. Initially, we identify potentially vulnerable libraries by systematically exploring the popular library categories on Maven~\cite{maven} and selecting libraries tagged with vulnerabilities in particular versions. To assess the efficacy of mitigations, we restrict our analysis to vulnerabilities accompanied by published exploits or those for which we develop exploits based on available information, such as vulnerability witness tests. Vulnerabilities lacking associated exploits are omitted from our dataset, resulting in the collection of \dx{94} vulnerabilities across \dx{48} libraries. To ensure the generalizability of the dataset, we include vulnerabilities with diverse reproducing behaviors. Our dataset covers \dx{35} CWEs, \dx{which collectively correspond to 3,421 historical Maven vulnerabilities, accounting for 58.09\% of the total.} Furthermore, the vulnerable libraries in our dataset span various technical domains, including  JSON libraries, logging frameworks, XML processing, compression libraries, and I/O utilities. \dx{Reproducing these vulnerabilities and cross-validating the reproduction outcomes against the expected behavior took the first two authors one month.}

\noindent \subsubsection{Vulnerable Functions.} After identifying vulnerabilities, we collect impacted functions for each vulnerability. A function is considered impacted if the project depends on a vulnerable library version, and the vulnerability can be replicated with specific input. Due to the limitations of GitHub's advanced search functionality, we implement two strategies to identify impacted functions: \ding{182} identifying projects that depend on a particular version of a library by examining the \textit{pom.xml} file and analyzing functions invoking the vulnerable library API $F_v$;  \ding{183} locating functions invoking $F_v$ and verifying the library versions.  \dx{Subsequently, we manually examine the candidate functions and relevant configuration files to confirm whether the vulnerabilities are exploitable in the target projects. We then attempt to build these projects by following their README and resolving the compilation issues encountered during the process. \xx{To make more projects executable in our evaluation environment, we performed minimal setup steps that do not affect the impacted downstream functions, including removing unrelated non-compilable modules and pinning unspecified libraries to the vulnerable versions.} Ultimately, we collected 474 impacted functions from successfully compiled projects. Collecting these affected functions and preparing the projects for successful compilation took the first two authors two months.}

%exploit test是什么，有什么特征
\noindent \subsubsection{Exploit Tests.} 
To determine whether a vulnerability is mitigated within a function, we construct exploit tests for each function, utilizing published exploits related to the vulnerability.  A vulnerability is considered mitigated if the exploit test fails to reproduce the behavior. When dealing with library vulnerabilities, an exploit test involves invoking functions in the project that cover $F_v$ and providing exploitable input to $F_v$ to reproduce the vulnerability. \xx{The assertion in each exploit test} is determined based on the observed behavior during reproduction. For different vulnerability manifestations, we adopt tailored verification strategies, including capturing the expected exception for \textit{Uncaught Exception}, using a thread watchdog with an execution time limit to detect runtime timeouts caused by \textit{Resource Exhaustion}, monitoring the target server to identify potential attacks in cases of \textit{Malicious Code Execution}, and comparing the observed return value against the expected behavior for \textit{Wrong Return Value}.
For each vulnerability, we gather exploit inputs from known exploits and manually modify them according to the impacted function to construct a downstream exploit test. We then verify the effectiveness of the exploit tests on vulnerable projects.

\subsubsection{Functionality Tests.}
\xx{To evaluate whether the generated mitigations preserve functionality, we collect functionality tests from three sources.} 
\ding{182} \textit{Constructed Tests.} We extract sample inputs and expected outputs of vulnerable API usages from library documentation and test suites. Based on these samples, we construct tests that use the expected outputs as test oracles to assess whether their behavior is preserved after mitigation. 
\ding{183} \textit{EvoSuite-Generated Tests.} To further increase the coverage of the class files containing impacted downstream functions, we generate additional tests using EvoSuite~\cite{EvoSuite}. 
\xx{\ding{184} \textit{Project Original Tests.} To better assess the impact of generated mitigations on real-world compatibility, we additionally execute the original test suites provided by the downstream projects in our evaluation dataset. 
For each downstream project with executable tests, we run the tests before and after applying the generated mitigation to identify newly introduced compatibility issues. In total, we execute 2,929 original tests, among which 1,961 passed before applying the mitigation. Overall, our functionality evaluation includes 6,377 valid tests, consisting of 473 downstream-behavior tests, 3,943 EvoSuite-generated tests, and 1,961 original tests.}

\subsection{Experiment Settings} 

\dx{ \subsubsection{Implementation Details.}
We utilize GPT-5.2 (specifically, the gpt-5.2-2025-12-11 version snapshot) across all stages of \appname, given its demonstrated strong performance on code-related tasks. The model is accessed via the official OpenAI API. We configure the decoding settings with a temperature of 0.5 and a top\_p of 1.0. Furthermore, we leverage the default maximum context window to process comprehensive context information, while the maximum output length (max\_tokens) is constrained to 4096 tokens.
We implement the agent workflow using \textit{autogen} (version 0.10.5), an open-source framework that supports agent orchestration and tool invocation. This framework allows \appname to coordinate context collection, mitigation generation, and validation within a unified workflow.
We use the same configuration for the LLM-based baselines to ensure a fair comparison.  We use JavaParser~\cite{javaparser} to obtain the AST of the target function.}

\dx{ \subsubsection{Interaction Budget.}
For mitigation context collection, we set the overall interaction budget to 30 rounds. Once this budget is exhausted, the context collection process terminates, and mitigation generation proceeds based on the context collected so far. In addition, each individual collection goal is allowed up to 5 rounds for satisfaction. If the goal remains unmet after 5 rounds, it is marked as unsatisfied. Finally, if a target location fails security effectiveness validation after five attempts, or no valid mitigation can be generated after five attempts, we mark that location as unmitigable in our evaluation. }

\subsection{Baselines and Evaluation Metrics}

\noindent \textbf{Baseline.} We choose \baselinename as our baseline ~\cite{huang2019rapid}, which neutralizes vulnerabilities within a binary by throwing a processable error code at the beginning of the vulnerable function to block the execution and prevent malicious attacks. As \baselinename is tailored for C++ projects, to conduct a comparative analysis between \appname and \baselinename, we assess the feasibility of substituting the vulnerable function with a handleable exception within the project.  

\dx{APR and AVR methods are also important baselines in our evaluation. We select one representative APR method, \thinkbold~\cite{Yin2024ThinkRepair}, and two representative AVR methods, \musselbold~\cite{Jiao2025MUSSEL} and \VulMasterbold~\cite{Zhou2024Vulmaster}, as our baselines.}

\begin{itemize}
    \item \dx{\textbf{\thinkbold} is an LLM-based automated program repair approach designed to address bugs that require program reasoning. It consists of a \emph{collection phase}, which automatically gathers chain-of-thought examples as repair knowledge, and a \emph{fixing phase}, which performs few-shot bug fixing with optional test feedback. Its key strength lies in leveraging automatically collected reasoning traces to improve repair quality.}

    \item \dx{\textbf{\musselbold} is a vulnerability repair framework based on LLMs. It adopts a multi-stage training pipeline, including supervised fine-tuning on vulnerability-repair data and self-reward feedback learning to iteratively improve patch generation. It is specifically designed to enhance the precision and efficiency of vulnerability patching across diverse CWEs.}

    \item \dx{\textbf{\VulMasterbold} is a Transformer-based neural vulnerability repair model that aims to generate repairs by jointly modeling vulnerable code, code structure, and expert knowledge from the CWE system. \VulMasterbold emphasizes structural understanding and the integration of vulnerability-specific expert knowledge.}
\end{itemize}

\dx{For \musselbold{} and \VulMasterbold{}, we use the released model weights provided in their replication packages. Following their repair setting, we mark the vulnerable API invocation line with \texttt{<vul-start>} and \texttt{<vul-end>} as the model input and then evaluate their generated patches in our downstream mitigation setting. For \thinkbold{}, we use the constructed exploit tests to validate its first-stage trace generation. To ensure a fair comparison, we configure \thinkbold with the same base model as \appname (GPT-5.2).}

\noindent \textbf{Evaluation Metrics.} We evaluate the performance of \appname by employing two metrics utilized in previous studies~\cite{Huang2016Neu, huang2019rapid}, namely security and functionality. Security evaluates the efficacy of the implemented mitigation. Consistent with Huang et al.~\cite{Huang2016Neu}, we create an exploit test for the impacted function. The mitigation is classified as a security enhancement if the exploit test fails to replicate the vulnerability. To determine the compatibility of \xx{a} generated mitigation, we construct a functionality test dataset in our data collection process for impacted functions to identify \dx{a} loss in functionality. The library vulnerability within the function is considered \textit{mitigated} if the exploit test fails to reproduce the vulnerability and is considered \textit{safely mitigated} \xx{if no new test failures occur} when executing the functionality test dataset.

\section{Results}

In this section, we conduct experiments to answer our RQs.

\subsection{RQ1: Effectiveness of \appnamebold}

\dx{\textbf{Overall Performance.} As shown in Table ~\ref{tab:performance_comparison}, \appname achieves an overall mitigation success rate of 82.91\% (393/474), substantially outperforming all baseline approaches.  Specifically, \appname achieves a 139.63\% relative improvement over the SOTA vulnerability mitigation baseline, which has a success rate of 34.60\% (164/474). This result demonstrates the stronger performance of \appname under our downstream exploit-based evaluation.}

\dx{\textbf{Analysis of Baseline Failures.} A critical observation from Table~\ref{tab:performance_comparison} is the poor performance of the APR and AVR tools, with the latter failing to generate compilable patches. We conducted a detailed analysis of their performance.}

\begin{itemize}[leftmargin=*]
    \item \dx{\textbf{Context and Cross-Language Incompatibility in AVR.} \VulMaster and \mussel are primarily designed to repair vulnerabilities at their root cause within the library source code. Their fine-tuning and alignment largely rely on CWE-specific vulnerability-fix pairs, many of which are dominated by C/C++ patches. When applied to downstream projects that provide only a single-line API call as context, they exhibit two critical failure modes. \ding{182} First, a \textit{context mismatch} problem arises. Under such a limited context, \VulMaster generates a semantically meaningless patch for CVE-2021-21341, i.e., \texttt{<vul-start> size, 0 <vul-end>}. \ding{183} Second, they suffer from \textit{cross-language \xx{incompatibility} issues}. For example, when attempting to mitigate the Java deserialization vulnerability CVE-2022-38751, \mussel incorrectly produces C/C++ style repair patterns in a Java context using the pointer access operator (\texttt{->}), returning the C-style \texttt{NULL} macro, and invoking a non-existent function (\texttt{yaml\_free()}).}
    
   \item \dx{\textbf{Task Mismatch in APR.} APR methods (e.g., \think) also struggle in this task, achieving an effectiveness rate of only 16.88\% (80/474). We attribute this to the mismatch in task formulation. \think explicitly prompts the underlying model to `fix this fault' using local context with few-shot cases. However, the downstream API call itself is typically logically correct, while the vulnerability is reproduced within the library.  Consequently, this misaligned prompt causes a breakdown in the two-phase reasoning process of \think. In its initial generation phase, \think only formulated a valid trace for 13 functions, resulting in its low effectiveness.}
\end{itemize}

\begin{table}[htbp]
  \centering
  \caption{Performance Comparison of Different Vulnerability Handling Approaches}
  \label{tab:performance_comparison}
  \resizebox{\linewidth}{!}{
  \begin{tabular}{lccccc}
    \toprule
    \multirow{2.5}{*}{\textbf{Approach}} & \multicolumn{2}{c}{\textbf{Mitigation}} & \multicolumn{1}{c}{\textbf{APR}} & \multicolumn{2}{c}{\textbf{AVR}} \\
    \cmidrule(lr){2-3} \cmidrule(lr){4-4} \cmidrule(lr){5-6}
    & \appnamebold & \makecell{\baselinenamebold ~\cite{huang2019rapid}} & \makecell{\thinkbold ~\cite{Yin2024ThinkRepair}} & \makecell{\musselbold~\cite{Jiao2025MUSSEL} } & \makecell{\VulMasterbold ~\cite{Zhou2024Vulmaster}} \\
    \midrule
    \textbf{Effectiveness} & 82.91\% (393/474) & 34.60\% (164/474) & 16.88\% (80/474) & No valid patch & No valid patch \\
    \bottomrule
  \end{tabular}
  }
\end{table}

\dx{\textbf{Generalizability Across CWEs.} To evaluate the generalizability of \appname, we analyzed its performance across different CWEs, as detailed in Table ~\ref{tab:cwe_results}. \appname demonstrates generally strong performance across the evaluated CWE categories, particularly in mitigating memory and resource management issues, achieving a 94.23\% success rate for Out-of-bounds Write (CWE-787) and 100\% for Uncontrolled Resource Consumption (CWE-400). While maintaining a high overall success rate, \appname encounters challenges in specific complex scenarios, such as Improper Input Validation (CWE-20, 40.00\%) and Code Injection (CWE-94, 57.14\%). We attribute this to the fact that mitigating these vulnerabilities often requires a deep, context-specific understanding of the vulnerability trigger logic, which may not be fully captured by vulnerability reports or reproducing behaviors. Nevertheless, \appname still consistently outperforms baselines even within these categories.}

\begin{table*}[htbp]
\centering
\caption{Results of Our Effectiveness Evaluation w.r.t CWE Types (\#V. denotes the number of vulnerabilities)}
\label{tab:cwe_results}
\small
\resizebox{\textwidth}{!}{
\begin{tabular}{lccccccccccc}
\toprule
\multirow{2}{*}{\textbf{Metric}} & \textbf{CWE-502} & \textbf{CWE-787} & \textbf{CWE-611} & \textbf{CWE-121} & \textbf{CWE-434} & \textbf{CWE-770} & \textbf{CWE-400} & \textbf{CWE-20} & \textbf{CWE-94} & \textbf{CWE-22} & \textbf{OTHERS} \\
\cmidrule(lr){2-11} \cmidrule(lr){12-12}
& \#V.=10 & \#V.=11 & \#V.=7 & \#V.=5 & \#V.=4 & \#V.=6 & \#V.=2 & \#V.=3 & \#V.=4 & \#V.=4 & \#V.=38 \\
\midrule
\multirow{2}{*}{\appnamebold} 
& 85.58\% & 94.23\% & 80.00\% & 91.67\% & 89.66\% & 69.23\% & 100.00\% & 40.00\% & 57.14\% & 76.92\% & 82.26\% \\
& (89/104) & (49/52) & (36/45) & (33/36) & (26/29) & (18/26) & (16/16) & (6/15) & (8/14) & (10/13) & (102/124) \\

\midrule

\multirow{2}{*}{\thinkbold} 
& 42.31\% & 7.69\% & 35.56\% & 2.78\% & 6.90\% & 3.85\% & 0.00\% & 0.00\% & 21.43\% & 0.00\% & 7.26\% \\
& (44/104) & (4/52) & (16/45) & (1/36) & (2/29) & (1/26) & (0/16) & (0/15) & (3/14) & (0/13) & (9/124) \\

\midrule

\multirow{2}{*}{\baselinenamebold} 
& 50.96\% & 30.77\% & 33.33\% & 2.78\% & 13.79\% & 53.85\% & 12.50\% & 20.00\% & 35.71\% & 30.77\% & 37.90\% \\
& (53/104) & (16/52) & (15/45) & (1/36) & (4/29) & (14/26) & (2/16) & (3/15) & (5/14) & (4/13) & (47/124) \\

\bottomrule
\end{tabular}
}
\end{table*}

\begin{center}
    \resizebox{\linewidth}{!}{
\begin{tabular}{l!{\vrule width 1pt}p{0.9\columnwidth}}
    \makecell{{\LARGE \faLightbulbO}}  &\dx{\textbf{Answer to RQ1:}
\appname effectively mitigates third-party library vulnerabilities at the downstream project level, achieving an 82.91\% success rate.  It demonstrates generalizability across diverse vulnerability types, particularly in memory and resource management issues, while APR and AVR methods perform poorly in this scenario due to context mismatches, cross-language incompatibilities, and misaligned task formulations.}\\
\end{tabular}}
\end{center}

\subsection{RQ2: Key Designs of \appnamebold}
~\label{5.2}

The results of RQ1 demonstrate the overall effectiveness of \appname in mitigating library vulnerabilities in downstream projects. In RQ2, we further investigate the contribution of its key design components.   \dx{To achieve this, we conduct an ablation study along three key dimensions: \textbf{strategy composition}, \textbf{strategy selection}, and \textbf{context collection}. Based on these dimensions, we construct five degraded variants of \appname for comparison:}

\begin{itemize}[leftmargin=*]
    \item \dx{\textbf{$\text{V}_{\text{generated}}$ (Generated-strategies Only):} This variant removes CVE-level, CWE-level, and library-specific mitigation knowledge. It relies exclusively on the LLM's reasoning over the vulnerability description to construct mitigation strategies.}
    
    \item \dx{\textbf{$\text{V}_{\text{existing}}$ (Existing Strategies Only):} This variant removes the new strategy generation module, restricting the strategy pool entirely to existing mitigation knowledge collected from vulnerability reports, CWE entries, and historical mitigation records of the vulnerable library.}

    \item \dx{\textbf{$\text{V}_{\text{unfiltered}}$ (Unfiltered Strategies):} This variant removes the strategy selection module. Instead of filtering the strategy pool to identify the most applicable solutions, the LLM is directly provided with all constructed and retrieved strategies to guide the generation of the mitigation patch.}
    
    \item \dx{\textbf{$\text{V}_{\text{static-ctx}}$ (Static Context):} This variant removes the tool-augmented agentic context exploration process (e.g., file reading and tree traversal). Instead of dynamically gathering mitigation-relevant context, it is directly provided with the file containing the vulnerable API invocation as the sole context for mitigation generation.}
    
    \item \dx{\textbf{$\text{V}_{\text{LLM-only}}$ (Naive LLM Only):} This variant removes both strategy construction and context exploration. The LLM is only given the vulnerability description, the mitigation task, and the vulnerable file, and is asked to generate the mitigation patch end-to-end.}
\end{itemize}

Table~\ref{tab:cwe_ablation_results} reports the overall and per-CWE
mitigation results of \appname{} and its five variants.

\begin{table*}[htbp]
\centering
\caption{Results of Our Ablation Study w.r.t CWE Types (\#V. denotes the number of vulnerabilities)}
\label{tab:cwe_ablation_results}
\small
\resizebox{\textwidth}{!}{
\begin{tabular}{lcccccccccccc}
\toprule
\multirow{2}{*}{\textbf{Metric}} & \textbf{TOTAL} & \textbf{CWE-502} & \textbf{CWE-787} & \textbf{CWE-611} & \textbf{CWE-121} & \textbf{CWE-434} & \textbf{CWE-770} & \textbf{CWE-400} & \textbf{CWE-20} & \textbf{CWE-94} & \textbf{CWE-22} & \textbf{OTHERS} \\
\cmidrule(lr){2-12} \cmidrule(lr){13-13}
& \#V.=94 & \#V.=10 & \#V.=11 & \#V.=7 & \#V.=5 & \#V.=4 & \#V.=6 & \#V.=2 & \#V.=3 & \#V.=4 & \#V.=4 & \#V.=38 \\
\midrule
\multirow{2}{*}{\appnamebold} 
& 82.91\% & 85.58\% & 94.23\% & 80.00\% & 91.67\% & 89.66\% & 69.23\% & 100.00\% & 40.00\% & 57.14\% & 76.92\% & 82.26\% \\
& (393/474) & (89/104) & (49/52) & (36/45) & (33/36) & (26/29) & (18/26) & (16/16) & (6/15) & (8/14) & (10/13) & (102/124) \\

\midrule
\multirow{2}{*}{\textbf{$\text{V}_{\text{generated}}$}} 
& 73.42\% & 78.85\% & 88.46\% & 40.00\% & 97.22\% & 75.86\% & 61.54\% & 100.00\% & 20.00\% & 28.57\% & 92.31\% & 75.81\% \\
& (348/474) & (82/104) & (46/52) & (18/45) & (35/36) & (22/29) & (16/26) & (16/16) & (3/15) & (4/14) & (12/13) & (94/124) \\

\midrule
\multirow{2}{*}{\textbf{$\text{V}_{\text{existing}}$}} 
& 73.00\% & 80.77\% & 55.77\% & 86.67\% & 41.67\% & 96.55\% & 76.92\% & 81.25\% & 26.67\% & 50.00\% & 61.54\% & 79.84\% \\
& (346/474) & (84/104) & (29/52) & (39/45) & (15/36) & (28/29) & (20/26) & (13/16) & (4/15) & (7/14) & (8/13) & (99/124) \\

\midrule
\multirow{2}{*}{\textbf{$\text{V}_{\text{unfiltered}}$}} 
& 74.68\% & 82.69\% & 65.38\% & 82.22\% & 97.22\% & 93.10\% & 46.15\% & 100.00\% & 20.00\% & 28.57\% & 69.23\% & 73.39\% \\
& (354/474) & (86/104) & (34/52) & (37/45) & (35/36) & (27/29) & (12/26) & (16/16) & (3/15) & (4/14) & (9/13) & (91/124) \\

\midrule
\multirow{2}{*}{\textbf{$\text{V}_{\text{static-ctx}}$}} 
& 63.92\% & 69.23\% & 71.15\% & 71.11\% & 44.44\% & 75.86\% & 53.85\% & 87.50\% & 26.67\% & 28.57\% & 38.46\% & 66.94\% \\
& (303/474) & (72/104) & (37/52) & (32/45) & (16/36) & (22/29) & (14/26) & (14/16) & (4/15) & (4/14) & (5/13) & (83/124) \\

\midrule
\multirow{2}{*}{\textbf{$\text{V}_{\text{LLM-only}}$}} 
& 36.92\% & 48.08\% & 34.62\% & 28.89\% & 8.33\% & 75.86\% & 26.92\% & 68.75\% & 0.00\% & 7.14\% & 23.08\% & 37.90\% \\
& (175/474) & (50/104) & (18/52) & (13/45) & (3/36) & (22/29) & (7/26) & (11/16) & (0/15) & (1/14) & (3/13) & (47/124) \\

\bottomrule
\end{tabular}
}
\end{table*}

\dx{\textbf{Limitations of LLMs in Mitigation Generation.} We first examine the performance of \textbf{$\text{V}_{\text{LLM-only}}$}, which represents the direct use of the base LLM without any of our designed workflows. Despite using an advanced LLM, this variant achieves a mitigation rate of only 36.92\% (175/474). This substantial performance gap represents a 55.47\% relative decrease compared with \appname, strongly validating our core motivation that naively prompting an LLM with a vulnerability description and the file invoking the vulnerable API is highly inadequate for generating secure library mitigations in downstream projects.}

\dx{\textbf{The Effectiveness of Strategy Composition.} In terms of overall mitigation performance, both \textbf{$\text{V}_{\text{generated}}$} (73.42\%) and \textbf{$\text{V}_{\text{existing}}$} (73.00\%) perform worse than \appnamebold (82.91\%). While their overall effectiveness is similar, their performance is complementary across vulnerability types. For example, \textbf{$\text{V}_{\text{generated}}$} excels at memory-related vulnerabilities, achieving 97.22\% on CWE-121 (Stack-based Buffer Overflow), whereas \textbf{$\text{V}_{\text{existing}}$} performs much worse in this category (41.67\%). Conversely, for vulnerabilities whose mitigation relies heavily on configuration, such as CWE-611 (Improper Restriction of XML External Entity Reference), \textbf{$\text{V}_{\text{existing}}$} substantially outperforms \textbf{$\text{V}_{\text{generated}}$} (86.67\% vs. 40.00\%) by effectively leveraging established XML parsing workarounds. By selecting between these strategy sources, \appname achieves strong performance across categories, achieving relative improvements of 12.93\% and 13.58\% over \textbf{$\text{V}_{\text{generated}}$} and \textbf{$\text{V}_{\text{existing}}$}, respectively, demonstrating the effectiveness of our strategy composition design across diverse CWEs.}

\dx{\textbf{The Effectiveness of Strategy Selection.} We further examine \textbf{$\text{V}_{\text{unfiltered}}$}, and find that it achieves an overall mitigation rate of 74.68\% (354/474), representing a 9.92\% relative decrease compared with \appnamebold (82.91\%, 393/474). This result shows that directly exposing the LLM to the entire strategy pool is less effective than explicitly selecting the most suitable strategy for the current attempt. A likely reason is that unfiltered strategies introduce irrelevant or less compatible mitigation options, making it harder for the LLM to focus on the most appropriate solution under the current repository context. In contrast, our strategy selection design helps narrow the decision space, thereby improving the overall mitigation effectiveness.}

\dx{\textbf{The Effectiveness of Dynamic Context Exploration.} The \textbf{context collection} dimension further highlights the necessity of our tool-augmented context collection. \textbf{$\text{V}_{\text{static-ctx}}$}, which restricts the mitigation context, achieves only 63.92\% (303/474), representing a 22.90\% relative decrease compared with \appname. This limitation is especially problematic for vulnerabilities whose mitigation depends on cross-file context. For instance, for CWE-22 (Path Traversal), the success rate drops from 76.92\% with \appname{} to 38.46\% with \textbf{$\text{V}_{\text{static-ctx}}$}.
This indicates that generating valid mitigations often requires exploring cross-file context, which \appname explicitly facilitates through its tree traversal, grep search, and file reading capabilities.}

\begin{center}
    \resizebox{\linewidth}{!}{
    \begin{tabular}{l!{\vrule width 1pt}p{0.9\columnwidth}}
        \makecell{{\LARGE \faLightbulbO}}  &\dx{\textbf{Answer to RQ2:}
        The performance of \appname relies heavily on its key design components. Its strategy composition design effectively integrates the generated mitigation strategies with retrieved existing mitigation knowledge, overcoming the limitations of either strategy in isolation. Moreover, the dynamic strategy selection and context exploration are essential, as restricting mitigation generation to static local context weakens the ability to handle vulnerabilities whose mitigation depends on broader project context.}\\
    \end{tabular}}
\end{center}

\subsection{RQ3: Functionality Impact of \appnamebold}

\dx{Table~\ref{tab:test_failure_summary} shows that \appnamebold preserves functionality for the vast majority of evaluated tests. Across \xx{6,377} functionality tests, \xx{6,272} tests remain consistent after mitigation, accounting for \xx{98.35\%} of the total. \xx{Notably, all \xx{1,961} original tests that passed before mitigation continue to pass after applying the generated mitigations, showing that no newly introduced failures are observed in these executable downstream test suites.} The remaining 105 inconsistent tests come from our constructed and EvoSuite-generated tests, indicating that the generated mitigations introduce limited observed functional impact.
Among these inconsistencies, behavior change is the largest category \xx{(56 tests, 0.88\%)}, suggesting that some mitigations alter the original behavior. Over-strengthened mitigations account for 33 tests \xx{(0.52\%)}, where the introduced checks are stricter than necessary. Environmental artifacts are the smallest category \xx{(16 tests, 0.25\%)}, mainly caused by differences in the runtime environment, sandbox restrictions, or external dependencies rather than the mitigation logic itself.}

\begin{itemize}[leftmargin=1.2em]
    \item \dx{\textbf{Behavior Change.}
    An example of behavior change after mitigation is \textit{CVE-2023-34620}. In this case, the mitigated function still processes the original benign input, but changes its output representation. The functional test expects the parsed value to be \textit{alice}, whereas the mitigated version returns \textit{``alice''}. This kind of failure does not indicate that the input is rejected or causes a severe runtime error, but changes the output of an originally supported input.}

    \item \dx{\textbf{Over-strengthened.} A representative example of over-strengthened mitigation is \textit{CVE-2022-25857}. Here, the mitigation introduces a stricter YAML validation rule, causing the parser to reject inputs with the error \textit{``YAML line too long''}. This failure indicates that the generated defense is stronger than necessary.}

    \item \dx{\textbf{Execution Environment Issue.}
    A representative example is \textit{CVE-2022-21126}. The original implementation uses the library utility \textit{IOUtil.createTempDir}, whereas the mitigated version switches to a direct call to \textit{Files.createDirectories} to enforce safer temporary-directory creation. Under the EvoSuite sandbox, this direct directory write is blocked by file-permission restrictions, causing the test to fail.}
\end{itemize}

\dx{\textbf{Different categories of libraries tend to exhibit different patterns of functional impact after mitigation.} Among the inconsistent cases, \textit{Behavior Change} is most commonly observed in JSON/HJSON/HTML processing (39.28\%). \textit{Over-strengthened} is mainly concentrated in YAML, deserialization, and content-sanitization libraries (42.42\%), followed by compressed/archive and binary-format libraries (30.30\%). \textit{Execution Environment Issue} is most frequently associated with libraries that depend on temporary files or temporary directories (43.75\%).}

\begin{table}[htbp]
    \centering
    \caption{Summary of Functionality Test Results and Failure Reasons}
    \label{tab:test_failure_summary}
    \renewcommand{\arraystretch}{1.3}
    \resizebox{\columnwidth}{!}{
    \begin{tabular}{l c c c c c}
        \toprule
         & \multirow{2}{*}{\textbf{Total}} & \multirow{2}{*}{\textbf{Consistent Tests}} & \multicolumn{3}{c}{\textbf{Inconsistent Tests}} \\
        \cmidrule(lr){4-6}
        & & & \textbf{Behavior Change} & \textbf{Over-strengthened} & \textbf{Environment} \\
        \midrule

         \textbf{\xx{Project Original}} 
        & \xx{1,961} 
        & \xx{1,961 (100.00\%)} 
        & \xx{0 (0.00\%)} 
        & \xx{0 (0.00\%)} 
        & \xx{0 (0.00\%)} \\

        \textbf{Constructed} 
        & 473 
        & 436 (92.18\%) 
        & 18 (3.81\%) 
        & 16 (3.38\%) 
        & 3 (0.63\%) \\
        
        \textbf{EvoSuite-Generated} 
        & 3,943 
        & 3,875 (98.28\%) 
        & 38 (0.96\%) 
        & 17 (0.43\%) 
        & 13 (0.33\%) \\

        \midrule
        \textbf{\xx{Overall}} 
        & \xx{6,377 }
        & \xx{6,272 (98.35\%)} 
        & \xx{56 (0.88\%)} 
        & \xx{33 (0.52\%)} 
        & \xx{16 (0.25\%)} \\
        \bottomrule
    \end{tabular}
    }
\end{table}

\begin{center}
    \resizebox{\linewidth}{!}{
    \begin{tabular}{l!{\vrule width 1pt}p{0.9\columnwidth}}
        \makecell{{\LARGE \faLightbulbO}}  &\dx{\textbf{Answer to RQ3:}
        \appname introduces only a limited functionality impact. Across \xx{6,377} functionality tests, \xx{6,272 tests (98.35\%)} remain consistent after mitigation, showing that \xx{\appname introduces limited compatibility loss to downstream projects}.}\\
    \end{tabular}}
\end{center}

\subsection{Generalization to Unseen Vulnerabilities}

\dx{To examine whether \appname can generalize to newly disclosed vulnerabilities, we replace the model with GPT-5.1, whose knowledge cutoff is September 30, 2024~\cite{GPT5.1}. To obtain downstream exploits disclosed after this cutoff date, we construct an evaluation set based on the library exploit dataset built by Chen et al.~\cite{chen2025diffploit}, which contains five exploits for post-cutoff library vulnerabilities. \xx{We further reproduce two additional post-cutoff vulnerabilities, resulting in seven post-cutoff vulnerabilities in total.} Following the same procedure used in our data collection, we further identify downstream projects affected by these upstream vulnerabilities. As shown in Table~\ref{tab:unseen-vulns}, we finally collect \xx{19} vulnerable call sites from \xx{17} downstream projects that invoke the vulnerable functions.}

\begin{table*}[htbp]
\centering
\caption{Mitigating unseen vulnerabilities disclosed after the model cutoff date.}
\label{tab:unseen-vulns}
\small
\resizebox{\textwidth}{!}{
\begin{tabular}{ccccl}
\toprule
\textbf{CVE} & \textbf{Disclosed Date} & \textbf{Downstream Project} & \textbf{Mitigation State} & \textbf{Mitigation Strategy} \\
\midrule
CVE-2025-59340 & 2025-09-17 & jinjava-maven-plugin & 1/1 & Template allowlist \\
\midrule
CVE-2025-58056 & 2025-09-03 & netty4-http-server & 1/1 & Strict chunk validation \\
\midrule

\multirow{5}{*}{CVE-2025-53864} 
& \multirow{5}{*}{2025-07-10} 
&  sd-jwt & 1/1 & JWT payload depth check \\
& & \xx{rancher/cattle} & \xx{0/1} & \xx{JWT payload depth check} \\
& & \xx{xxl-tool} & \xx{1/1} & \xx{JWT payload depth check} \\
& & \xx{mangooio} & \xx{1/1} & \xx{JWT payload depth check} \\
& & \xx{XunLuAdmin} & \xx{1/1} & \xx{JWT parse guard} \\
\midrule

\multirow{7}{*}{\xx{CVE-2025-52999}} 
& \multirow{7}{*}{2025-06-25} 
& ResttemplateDemo & 3/3 & Safe Jackson wrapper \\
& & \xx{best-pay-sdk} & \xx{1/1} & \xx{Bounded Jackson parsing} \\
& & \xx{MarkovNetwork} & \xx{1/1} & \xx{Jackson parse isolation} \\
& & \xx{Warcraft-Tools} & \xx{0/1} & \xx{Jackson input clamp} \\
& & \xx{aria2-api} & \xx{1/1} & \xx{Jackson input clamp} \\
& & \xx{CorporateGenealogy} & \xx{1/1} & \xx{Safe Jackson wrapper} \\
& & \xx{nju-lib-downloader} & \xx{1/1} & \xx{Jackson parse isolation} \\
\midrule

CVE-2025-48976 & 2025-06-16 & Wildbook & 1/1 & Multipart pre-validation \\
\midrule
\xx{CVE-2025-35036} & \xx{2025-06-03}  & \xx{titus-control-plane} & \xx{1/1} & \xx{Disable EL interpolation} \\
\midrule

\xx{CVE-2025-22228} & \xx{2025-03-20} & \xx{NanoMaven} & \xx{1/1} & \xx{Password length guard} \\

\midrule
\textbf{Total} & -- & \textbf{17 projects} & \textbf{17/19} & -- \\

\bottomrule
\end{tabular}
}
\end{table*}

\dx{As shown in Table~\ref{tab:unseen-vulns}, \appname successfully mitigates \textit{17 out of 19 vulnerable call sites} considered in this evaluation across the \textit{17} downstream projects, suggesting that \xx{\appname can derive mitigation strategies for several newly disclosed vulnerabilities}. For example, for CVE-2025-52999, \appname mitigates three affected call sites in \textit{ResttemplateDemo} by introducing a safe wrapper around \textit{ObjectMapper.readValue} that performs a bounded depth pre-scan before deserialization. For CVE-2025-53864, it prevents stack overflow in \textit{sd-jwt} by replacing unbounded disclosure JSON parsing with depth-bounded parsing. For CVE-2025-58056, \appname inserts a strict pre-decoder validation gate that rejects LF-only chunk-size terminators before the request reaches the vulnerable HTTP decoder. For CVE-2025-59340, it hardens \textit{jinjava-maven-plugin} by validating templates before rendering and restricting them to a safe subset of expressions. These results indicate that \appname can still derive effective mitigation strategies for newly disclosed vulnerabilities whose details are beyond the knowledge cutoff by combining mitigation knowledge with repository-level context.}

\begin{center}
    \resizebox{\linewidth}{!}{
    \begin{tabular}{l!{\vrule width 1pt}p{0.9\columnwidth}}
        \makecell{{\LARGE \faLightbulbO}}  &\dx{\textbf{Answer to RQ4:}
        \xx{\appname remains effective on several post-cutoff vulnerabilities in our evaluation dataset.} Using GPT-5.1 with a knowledge cutoff of September 30, 2024, \appname successfully mitigates \xx{most} vulnerable call sites across \xx{17} downstream projects for vulnerabilities disclosed after the cutoff date. These results suggest that \appname does not rely solely on memorized knowledge to generate a mitigation.}\\
    \end{tabular}}
\end{center}
\section{Discussion}

\dx{In this section, we first present how \appname addresses corner cases. We then examine its cross-language generalizability, provide a manual inspection of the correctness of the generated mitigations, and finally discuss the threats to validity.}

\label{sec: discussion}

\subsection{Applicability of \appnamebold to Edge Cases}

\subsubsection{\xx{Beyond Single-Path Exploit Validation.}}

\xx{ Our mitigation effectiveness evaluation relies on constructed exploits to determine whether a generated mitigation blocks the vulnerable behavior. However, in real-world scenarios, the same upstream vulnerability may be triggered through multiple vulnerable library APIs. Moreover, the exploit payload may have semantically equivalent variants that differ from our constructed payloads while still reproducing the same vulnerability behavior. Therefore, we further evaluate whether the mitigations generated by \appname remain effective under these conditions: \ding{182} multiple vulnerable library APIs and \ding{183} semantic variants of the original exploit payload.}

\xx{First, we examine vulnerabilities that can be triggered through multiple vulnerable library APIs. For example, CVE-2018-1000632 in \textit{dom4j} can be triggered through both \textit{Element\#addElement(String)} and \textit{Element\#addAttribute(String,String)}. For CVE-2022-25857 and CVE-2022-42003, different payload forms or input representations may invoke different overloaded parsing APIs, all of which can reach the vulnerable behavior. We therefore evaluate whether the mitigations generated by \appname remain effective across these vulnerable APIs, rather than only on the API invoked by the original exploit.  As illustrated in Table~\ref{tab:multiple-vulnerable-apis}, the mitigation column reports the number of successfully mitigated downstream usages over the total vulnerable usages for each corresponding API, which shows that \appname can mitigate vulnerabilities with multiple trigger paths.}

\begin{table*}[t]
\centering
\caption{\xx{Examples of vulnerabilities with multiple vulnerable library APIs.}}
\label{tab:multiple-vulnerable-apis}
\small
\begin{tabular}{l l l c}
\toprule
\xx{\textbf{CVE}} & \xx{\textbf{Library}} & \xx{\textbf{Vulnerable APIs}} & \xx{\textbf{Mitigation}} \\
\midrule

\xx{CVE-2018-1000632} 
& \xx{dom4j}
& \makecell[l]{\xx{Element\#addElement(String)}\\ \xx{Element\#addAttribute(String,String)}}
& \makecell[c]{\xx{1/2}\\ \xx{1/1}}\\
\midrule

\xx{CVE-2022-25857} 
& \xx{SnakeYAML}
& \makecell[l]{\xx{Yaml\#load(String)}\\ \xx{Yaml\#load(InputStream)}\\ \xx{Yaml\#load(Reader)}}
& \makecell[c]{\xx{2/2}\\ \xx{3/4}\\ \xx{1/1}} \\
\midrule

\xx{CVE-2022-42003} 
& \xx{jackson-databind}
& \makecell[l]{\xx{ObjectMapper\#readValue(String, Class<T>)}\\ \xx{ObjectMapper\#readValue(String, JavaType)}\\ \xx{ObjectMapper\#readValue(JsonParser, JavaType)}}
& \makecell[c]{\xx{8/8}\\ \xx{3/3}\\ \xx{1/1}} \\

\bottomrule
\end{tabular}
\end{table*}

\xx{Second, we examine semantic variants of the original exploit payloads. We modify the payload in our dataset while preserving the same vulnerability reproduction behavior. For CVE-2022-42889, the original payload \textit{\$\{script:javascript:195 + 324\}} triggers script execution. We further construct variants that exercise other lookup behaviors, including DNS lookup and URL lookup, to test whether the mitigation blocks the broader exploitation semantics rather than only the original script-execution payload. For CVE-2022-42003, we vary the JSON target type, such as changing it to \textit{Object.class}, and adjust the recursive nesting depth, thereby examining whether the generated mitigation can still prevent the same deep-nesting vulnerability behavior under different payload structures and parsing contexts. \appname successfully mitigates these semantic variants, suggesting that the generated mitigations are not merely exploit-specific checks for the original payloads. Instead, they can generalize to broader payload semantics.}

\subsubsection{\xx{Less common vulnerability types.}}
\dx{Rather than being constrained by specific vulnerability types mentioned in our preliminary study, \appname dynamically synthesizes mitigation knowledge through three complementary sources: \ding{182} CWE-level guidance, \ding{183} library-specific mitigation references, and \ding{184} LLM-driven strategy reasoning over the vulnerability description to infer downstream-applicable solutions. This design allows \appname to generalize across diverse vulnerability types and generate mitigations even when no existing remediation strategy is available.}

\dx{For example, consider CVE-2023-49250, in which the vulnerable library fails to verify certificates for outgoing HTTPS connections, thereby enabling \textbf{man-in-the-middle attacks}. To mitigate the vulnerability without upstream patches, the \textit{Strategy Collection} module first gathers structural mitigation knowledge from the CWE knowledge base, such as certificate pinning and proxy-based enforcement. Based on the CVE description, it further infers a downstream-oriented strategy, \textit{Downstream HTTPS Client Hardening Wrapper}, which is intended to mitigate the vulnerable behavior by enforcing strict TLS validation semantics. The \textit{Mitigation Module} then selects this strategy for execution and synthesizes a \texttt{SafeHttpClient} class, which preserves certificate-chain and hostname verification in outbound HTTPS communication. In this way, \appname replaces the vulnerable function with a hardened downstream-controlled implementation, thereby mitigating the risk exposed by utilizing the vulnerable library.}

\dx{Similarly, vulnerabilities causing timing attacks can also be mitigated by \appname. \xx{For instance, CVE-2017-9735 arises because the vulnerable library performs an unsafe password comparison, thereby exposing authentication operations to timing attacks.} Specifically, no explicit mitigation strategy is disclosed for this CVE, and the available CWE-level mitigations are not directly actionable for downstream repair, as they mainly focus on architectural privilege separation or restricting error-message disclosure rather than fixing the unsafe comparison logic itself. In this setting, \appname infers a mitigation strategy directly from the vulnerability description and replaces the original timing-sensitive comparison with a constant-time comparison, thereby reducing the risk that an attacker can infer the password from timing differences observed during authentication failures.}

\dx{Together, these examples show that the broad applicability of \appname comes from combining heterogeneous knowledge sources with generated strategies, rather than being limited to specific types of vulnerabilities. More importantly, this design allows \appname to infer and apply practical downstream mitigations even for vulnerabilities that lack existing remediation guidance. As a result, this design improves the applicability of \appname in real-world mitigation scenarios.}

\subsection{Repository-Level Mitigation with Automated Context Collection}

\dx{Certain vulnerabilities (e.g., CVE-2023-43643 and CVE-2024-23635) are not mitigated by modifying vulnerable API invocations or the instantiation of vulnerable classes directly, but by adjusting the repository-level security policy that enables the vulnerable behavior. In such cases, \appname must first locate the relevant policy file, inspect whether vulnerability-triggering settings (e.g., enabling \texttt{preserveComments} together with permissive tag policies in AntiSamy) are present, and then update the corresponding configuration to disable the risky behavior.}

\dx{In the case of CVE-2024-23635, \appname first analyzes the project file structure to locate the AntiSamy policy file \texttt{antisamy-html.xml} in the repository. After reading the file, it determines that effective mitigation requires the \texttt{preserveComments} directive to remain disabled. Based on this repository-level context, \appname then synthesizes the corresponding configuration update by inserting the directive \texttt{<directive name="preserveComments" value="false"/>}, together with an explanation that this setting is intended to mitigate CVE-2024-23635 (mXSS) for AntiSamy versions prior to 1.7.5 by preventing comments from being preserved in sanitized output. This example demonstrates that \appname can automatically perform repository-level mitigation by adjusting project configuration without manual intervention, thereby reducing the risk of potential malicious exploitation.}

\subsection{\xx{Cross-Language Evaluation of \appnamebold}}

\dx{We evaluate \appname on our Java-based downstream vulnerability dataset. \xx{To further explore whether \appname can be applied beyond the Java ecosystem}, we conduct an additional evaluation on C/C++ downstream projects. For this purpose, we adapt the dataset of ChainFuzz~\cite{Deng2025chainfuzz}, the largest public benchmark that provides exploit tests for reproducing upstream vulnerabilities in downstream projects. The ChainFuzz dataset contains 15 cross-library CVE cases. Due to the difficulty of reconstructing historical C/C++ projects and reproducing the corresponding vulnerabilities, we successfully established exploit tests for six downstream CVE vulnerabilities, which are illustrated in Table~\ref{tab:chainfuzz-reproduction}. These cases cover three widely used real-world C/C++ downstream projects: MuPDF, libheif, and libtiff.}

\begin{table}[t]
\centering
\caption{Cross-language mitigation results for C/C++ vulnerabilities.}
\label{tab:chainfuzz-reproduction}
\small
\begin{tabular}{lclc}
\toprule
\textbf{Project} & \textbf{CVE} & \textbf{Manifestation} & \textbf{Mitigation} \\
\midrule

\multirow{3}{*}{MuPDF} 
& CVE-2016-10506 & Division-by-zero leading to application crash (DoS). & $\checkmark$ \\
& CVE-2019-6988 & Excessive memory allocation. & $\checkmark$ \\
& CVE-2023-39328 & Application crash. & $\checkmark$ \\
\midrule

\multirow{2}{*}{libheif} 
& CVE-2023-24751 & NULL pointer dereference leading to denial of service. & $\checkmark$ \\
& CVE-2023-25221 & Heap buffer overflow. & $\checkmark$ \\
\midrule

libtiff 
& CVE-2021-29390 & Heap-based buffer over-read. & $\checkmark$ \\
\bottomrule
\end{tabular}
\end{table}

\dx{To adapt \appname to these C/C++ targets, the only necessary adaptation lies in the Verification Module, where we simply replace the Java-specific build commands (e.g., \textit{mvn compile}) with the corresponding C/C++ compilation commands (e.g., \textit{make} or \textit{cmake}) used by these downstream projects. Under this minimal adaptation, \appname successfully generates and applies actionable downstream mitigations that effectively block the vulnerability triggers in all 6 vulnerabilities. \xx{This result suggests that, after adapting the compilation and verification commands, \appname can also mitigate vulnerability triggers in C/C++ downstream projects. However, its effectiveness across more ecosystems and on a larger-scale dataset remains to be validated, which we leave as future work.}}

\dx{To illustrate this cross-language capability, consider the mitigation generated for CVE-2021-29390 in \textit{libtiff}, which originates from a heap-based buffer over-read in the \textit{libjpeg-turbo} library. Reproducing this vulnerability requires a progressive JPEG stream. The mitigation generated by \appname works by rejecting a dangerous class of inputs before decompression starts, thereby preventing the vulnerable upstream JPEG smoothing logic from being executed in the downstream context. Concretely, \appname inserts a check after \textit{TIFFjpeg\_read\_header()} but before \textit{TIFFjpeg\_start\_decompress()}, and aborts processing when a multi-scan progressive JPEG is detected. To align with common C/C++ development practice, the restriction is encapsulated in a compile-time setting to ensure compatibility. This example shows that \appname can generate mitigations beyond the Java ecosystem.}

% dataset:

% mupdf
% CVE-2019-6988: 增加了time watch dog，超过两秒kill
% CVE-2023-39328: 策略为JPX image too large
% CVE-2022-43239： 关闭feature，导致libheif 失去 libde265 的 HEVC 解码能力，但是阻止了漏洞的复现
% CVE-2023-24751：开启了一个防护线程

% First, we discuss why the APR and AVR methods are not selected as our baseline. Then, we discuss threats to validity of \appname.

\subsection{Mitigation Reliability of \appnamebold}
% \subsection{Manual Inspection of Mitigation Efficacy and Functional Compatibility}

\subsubsection{\xx{Failure Case Analysis}}
\xx{To better understand the limitations of \appname, we manually inspect representative failed mitigation cases and summarize recurring failure patterns, which show that unsuccessful mitigations often arise from incomplete repository-context collection or incorrect understanding of the concrete exploit condition.}

\ding{182} \textit{\xx{Missed alternative entry points.}}
\xx{Some failures occur because the generated mitigation protects only one downstream invocation path while other entry points still reach the vulnerable API. In these cases, the selected mitigation strategy may be appropriate, but the repository-context collection is incomplete. For example, in CVE-2022-24196, \appname adds size and page-count checks to the \textit{MultipartFile} upload method, but the exploit test invokes another byte-array-based upload method that still directly constructs PdfReader and PdfDocument without the required resource guards. As a result, the malicious input can bypass the generated mitigation through an uncovered downstream entry point.}

\ding{183} \textit{\xx{Misidentified exploit trigger path.}}
\xx{Some failures occur because \appname misidentifies the key features of the exploit payload. As a result, the generated mitigation filters irrelevant payload patterns and fails to block the actual exploit. In these cases, the generated mitigation is applied to the wrong trigger form or an incomplete entry chain, so the actual exploit path remains unblocked.}
\xx{For example, in several Jackson default-typing vulnerabilities, \appname assumes that the exploit relies on explicit type indicators such as \textit{@type} or \textit{@class}, and thus generates filters for these fields. However, the actual payload uses a wrapper-array representation, where the gadget class name is encoded as the first array element. Since the key payload feature is the wrapper-array type encoding rather than explicit type fields, the generated filter is bypassed. Similarly, in the XStream vulnerabilities, \appname blocks several known gadget classes through a manually written denylist converter, but the actual payload relies on an unlisted RMI/JNDI gadget chain. }

\subsubsection{\xx{Mitigation Stability Analysis}}

\xx{To further analyze the stability and calibration of \appname, we conduct a proportional random sampling analysis. We sample 60 cases from our dataset in RQ1, including 50 originally successful cases and 10 originally failed cases. At a 90\% confidence level, this sample size yields an approximate margin of error of 10\%. For each sampled case, we rerun \appname three times, record the LLM judgment during Security Effectiveness Validation (Section \ref{security}), and execute the exploit test to obtain the mitigation effectiveness.}

\xx{Across the 180 repeated runs, 144 runs successfully block the exploit test after mitigation, while the exploit tests still trigger the vulnerabilities in the remaining 36 runs. The rerun success rate is therefore 80.00\%, which is close to the original full-dataset mitigation rate of 82.91\%. At the case level, 48 out of the 60 sampled cases have consistent final mitigation status across all three repeated runs, while the remaining 12 cases show inconsistent status. These results suggest that \appname achieves relatively stable mitigation effectiveness in most sampled cases.}

\xx{To assess validation agreement, we compare the LLM judgment in  Security Effectiveness Validation with the exploit validation results. Among the 180 runs, 134 runs are judged as mitigated by \appname and also pass the exploit test, while 8 runs are judged as ineffective and also fail the exploit validation.  Overall, the LLM judgment agrees with the exploit in 142 out of 180 runs, yielding an agreement rate of 78.89\%. These results show that the LLM-guided validation is generally aligned with the final result. However, it still overestimates security effectiveness in some cases, where the mitigation is judged as effective despite failing the exploit validation. Improving this calibration remains an important direction for future work.}

% 我们进行了随机抽样，按照比例抽取了50个成功缓解的case以及10个失败缓解的case进行了3次重复实验（这里是否需要计算置信？）。我们记录了缓解中模型选择的策略，以及Security Effectiveness Validation中的feedback。

% 首先是对于stability，在这180个运行中，通过exploit测试进行验证，有144个运行成功缓解利用测试，36个运行仍在缓解后会被利用测试触发。这个结果与我们原本的的成功率接近，证明了\appname在成功生成缓解的任务上具有稳定性。在所有的60个case中，仅有12个case出现了三次缓解之中结果不一致的情况，证明了\appname的缓解具有稳定性

% 然后是calibration上，180次运行中，134个case上，LLM判断为Security Effectiveness且通过 success，8个case上LLM判断fail 且 oracle failure，说明在大部分情况下，LLM的判断与实际一致。但是我们也发现有28个case中，LLM判断缓解成功，但是实际上没有通过oracle，说明仍存在一定的over calibration，需要作为我们的future work来改进。

\subsubsection{\xx{Manual Inspection of Mitigation Efficacy and Functional Compatibility}}
\dx{To provide additional evidence beyond test-based validation, we further conducted a manual inspection of a randomly sampled subset of the mitigated cases. Specifically, out of the 393 cases successfully mitigated by \appname, we randomly sampled 80 cases for manual analysis. This sample size yields a 95\% confidence level with a margin of error of approximately 10\%. The first two authors, each with more than three years of experience in analyzing vulnerabilities in open-source software, independently assessed each sampled case along two dimensions: \textit{mitigation effectiveness} and \textit{semantic preservation}. Each dimension is scored on a three-point scale (0/1/2). For \textit{mitigation effectiveness}, a score of 2 indicates that the downstream exploit path was credibly blocked, 1 indicates partial or uncertain mitigation, and 0 indicates no credible mitigation. For \textit{semantic preservation}, a score of 2 indicates no clear evidence of breaking intended benign behavior, 1 indicates some risk of compatibility or behavior change, and 0 indicates a clear semantic regression. When the two reviewers disagree, the final decision is made by a third annotator with more than five years of experience in analyzing vulnerabilities in open-source software. }

\dx{The manual inspection indicates that most automatically successful cases indeed achieve credible downstream mitigation. Under the final adjudicated labels, 68 of the 80 sampled cases (85.0\%) received a mitigation-effectiveness score of 2, nine cases (11.3\%) received a score of 1, and only 3 cases (3.8\%) received a score of 0. This yields an estimated full-mitigation rate of 85.0\% (95\% CI: 78.0\%--92.0\%) among the 393 cases labeled as successful by automated evaluation. For semantic preservation, the results are generally positive. Specifically, 52 of the 80 sampled cases (65.0\%) received a score of 2, while 27 cases (33.8\%) received a score of 1, and only one case (1.3\%) received a score of 0. These results indicate that the generated mitigations are generally compatible with the target projects. However,  some patches still exhibit compatibility impacts, such as narrowing the range of accepted inputs.}

\subsection{Threats to Validity}
\textbf{External Validity.} One potential factor that could impact external validity is the possibility that our experiment's dataset may not encompass all vulnerability types and trigger conditions. \dx{To overcome this, we follow previous studies and construct a dataset that covers 35 CWEs to confirm the generalizability of our method.} \xx{Another threat to external validity stems from mitigation-strategy collection, which relies on domain knowledge extracted from vulnerability reports and may be limited when such knowledge is unavailable.} \dx{To address this limitation, we extend our mitigation knowledge by incorporating the \textit{Potential Mitigations} provided in CWE reports.}

\xx{When evaluating original tests, we try our best to reproduce the original testing environments of the downstream projects. However, some projects could not be fully reproduced because their test suites depended on missing test helper classes after compilation or required additional external services, such as database configurations. For these projects, we could not reliably execute the original tests. Meanwhile, some tests already failed before applying the generated mitigation, and thus were excluded from our functionality evaluation. Therefore, our functionality evaluation may still miss compatibility losses that would only be exposed under such requirements.}

Furthermore, \appname focuses on vulnerabilities in Java, and its performance in other programming languages has \xx{not yet been validated at scale}. \dx{Nevertheless, as discussed in our cross-language analysis on C/C++, \appname can be transferred to other languages with only minimal engineering adaptation, indicating its potential for cross-language mitigation generation.}

\textbf{Internal Validity.} \dx{One threat to internal validity is whether the observed mitigation performance mainly stems from the capabilities of the underlying LLM, rather than from the design of \appname itself. To overcome this, we design an ablation study to demonstrate that the mitigation ability of \appname primarily comes from the combination of strategy preparation, strategy selection, and repository-aware adaptation, rather than from relying solely on the base model.}

\dx{A threat to internal validity is that generated mitigations may unintentionally affect the original functionality. However, our results show that such an impact is limited. The observed inconsistencies are concentrated in a small number of cases, mainly involving behavior changes on original inputs, over-strengthened validation, and execution-environment-related issues.}

Another threat to internal validity is that we manually construct exploit tests for our experimental evaluation. Though we compare reproducing behaviors with vulnerability reports, we need further verification on whether our exploit tests correspond to actual vulnerability exploits. 
\xx{Moreover, \appname relies on the LLM for strategy selection and mitigation generation, which means different runs may select different strategies or generate different patches for the same case. To mitigate this threat, we conduct a repeated-run analysis. The results show that most sampled cases have consistent mitigation results across repeated runs, suggesting that the mitigation effectiveness of \appname is relatively stable in our sampled evaluation.}

\textbf{\xx{Construct Validity.}}
\xx{One threat to construct validity lies in the baseline comparison. Some compared baselines are originally designed for different vulnerability-handling settings. Specifically, \baselinename targets mitigation for C/C++ projects, and we adapt its core idea by checking whether the vulnerable downstream usage can be guarded by a handleable exception. \think, \mussel, and \VulMaster are designed for APR or AVR, and we adapt them by providing the downstream vulnerable code context and evaluating their generated patches for downstream projects. However, these adaptations may not fully reflect the intended usage scenarios of these approaches. Therefore, the baseline results should be interpreted as reference comparisons under our setting.}

\xx{Another threat to construct validity is that exploit-based validation may not fully capture all possible trigger paths of a vulnerability. A single exploit test may cover only one vulnerable API usage or one concrete payload form. To mitigate this threat, we further discuss two extended validation settings: vulnerabilities that can be triggered through multiple vulnerable library APIs, and vulnerabilities whose payload semantics can be varied while preserving the same reproduction behavior. This analysis provides initial evidence that \appname can remain effective in these settings. However, we do not systematically expand the exploits for all vulnerabilities in our dataset, leading to potential threats to construct validity.}

% Although our vulnerability classification method covers over 90\% of vulnerability types in widely used libraries, some vulnerability types remain uncovered, such as man-in-the-middle attacks. 

\section{Related Work}

\textbf{Vulnerability Mitigation.} Recent studies on vulnerability mitigation focus on directly mitigating vulnerabilities in the vulnerable library by hardening binary code and instrumenting code in vulnerable functions to prevent vulnerability triggers~\cite{huang2019rapid, Huang2016Neu}. Huang et al. proposed a simple method, \baselinename~\cite{huang2019rapid},  to mitigate security vulnerabilities by preventing vulnerable code from being executed and applying existing error-handling code in vulnerable projects. However, \baselinename led to a rejection of input \dx{that} is unable to trigger the vulnerability, affecting the project functionality. Many approaches have been proposed to \dx{harden} the binary code of programs to protect the programs from malicious attacks. SmashGadgets \cite{Pappas2012Smash} randomized binary code to hinder ROP attacks to prevent program instructions from being used as attack gadgets. HPAC \cite{zeng2018codelesspatchingheapvulnerabilities} changed the behavior of heap management routines and encoded the calling context to specific heap vulnerabilities. However, these methods are \xx{designed to address specific types of vulnerabilities}. In contrast, our study mitigates library vulnerabilities in downstream projects when modifying the vulnerable library is infeasible. Moreover, \xx{\appname is applicable to a diverse range of vulnerability types and introduces limited observed test-level compatibility impact in our evaluation}. 

\textbf{Library Vulnerability Remediation.} To address library vulnerabilities, developers have proposed approaches that involve automatically updating dependencies to secure versions or generating patches based on domain knowledge related to bug fixes. Software composition analysis is widely applied to reveal library vulnerabilities by adapting various remediation strategies. Dependency-management bots~\cite{R1, kula2018developers} are widely used to update dependencies when a library vulnerability is discovered. However, due to concerns about updates and notification fatigue, only 32\% of pull requests initiated by dependency management bots were merged~\cite{Mirhosseini1}. To address concerns about the compatibility, CORAL~\cite{Zhang1} implemented global optimization to enhance security and compatibility, achieving a success rate of 98.67\%. Patch generation requires domain knowledge, such as existing patches. Several research studies have generated patches leveraging previous patches in hard forks~\cite{pan2024automating}, existing open-source patches~\cite{Long1}, and already fixed applications  ~\cite{Sidiroglou1, Sidiroglou2}.  Hu et al.~\cite{Hu1} patched vulnerabilities by incorporating code from correct versions. Duan et al.~\cite{mitigation} addressed library vulnerabilities by integrating vulnerability patches into the binary file. These studies focused on patching using existing patches or fixed releases. \appname mitigates the vulnerability in impacted projects before official patches to achieve a quick response.

\dx{\textbf{Automated Program Repair.} Prior surveys have summarized the growing adoption of deep learning and LLMs in APR~\cite{chen2025deep, Yang2025Survey}. Representative APR methods include SequenceR~\cite{chen2019sequencer}, CoCoNuT~\cite{lutellier2020coconut}, CURE~\cite{jiang2021cure}, and AlphaRepair~\cite{xia2022less}, which formulate repair as a code generation task. Beyond neural code generation, L2S~\cite{Xiong2022L2S} models program repair as a probabilistic program estimation task and repairs conditional statements through a search-based framework. More recent studies further demonstrate the potential of LLM-based repair. To reduce reliance on external execution feedback and rigid prompting, ThinkRepair~\cite{Yin2024ThinkRepair} introduces a self-directed automated program repair framework that enables LLMs to autonomously analyze failures and iteratively refine patches. To further bridge the semantic gap between bug symptoms and human expectations, PATCH~\cite{Zhang2025Patch} enhances LLM-based repair by incorporating programmer-intent guidance and simulating collaborative debugging behaviors. Moreover, the scope of automated repair is expanding beyond textual source code into the visual domain. Huang et al. ~\cite{huang2025seeingfixingcrossmodalreasoning} pioneer the use of multimodal LLMs for cross-modal reasoning in visual software issue repair, targeting complex UI/UX anomalies.}

\dx{\textbf{Automated Vulnerability Repair.} Early LLM-based approaches mainly formulate vulnerability repair as a sequence generation task~\cite{Michael2022Repair, Pearce2023Repair}. However, these methods suffer from several limitations, especially limited input context and low patch precision. To address these issues, VulMaster~\cite{Zhou2024Vulmaster} expands the input by incorporating the entire vulnerable function, structural information, and CWE-based expert knowledge. Building on this line of research, Jiao et al.~\cite{Jiao2025MUSSEL} proposed MUSSEL, which further aims to reduce the reliance on extensive sampling and enable one-shot vulnerability patching. Instead of depending only on standard likelihood-based generation, MUSSEL adopts a multi-stage framework inspired by human learning to better align the model with high-precision repair objectives.}

\section{Conclusion}
In this paper, we propose an approach named \appname for mitigating library vulnerabilities without necessitating available patches or source code of libraries. \dx{\appname operates through a four-stage iterative process, comprising multi-source strategy preparation, context-aware strategy selection, tool-augmented downstream context collection, and adaptive mitigation generation, to securely neutralize vulnerabilities while preserving downstream compatibility.} We compile a dataset encompassing \dx{94} vulnerabilities from recent years, linked with \dx{474} impacted functions alongside exploit tests and \xx{6,377} functionality tests.  The experimental results, including \dx{393} successfully mitigated impacted functions and \xx{6,272} passed functionality tests, demonstrate the effectiveness of \appname. Our approach surpasses the baseline in neutralizing vulnerabilities with limited functionality loss. All the key designs of \appname contribute to mitigation generation. We also \dx{conduct} an empirical study to explore the criteria for classifying library vulnerabilities based on reproducing behaviors. Furthermore, we aim to investigate the performance of other LLMs in the task of generating mitigations.

\section{Data Availability}
\dx{The dataset and the code of \appname are publicly available on our website to support future research endeavors~\cite{Package}.}

\section*{Acknowledgements}
This research is supported by Fundamental Research Funds for the Central Universities (No. 226-2025-00171). We also thank the anonymous reviewers for their insightful comments and suggestions.

\bibliographystyle{ACM-Reference-Format}
\bibliography{main}

\end{document}